\documentclass[twocolumn,neutral,epjc3]{svjour3mod}
\usepackage[T1]{fontenc}
\usepackage{graphicx}
\usepackage{mathptmx}% use Times fonts if available on your TeX system
\usepackage{flushend}
\usepackage{booktabs} 
\usepackage{multirow}%
\usepackage{amsmath,amssymb,amsfonts}%
\usepackage{mathrsfs}%
\usepackage{xcolor}%
\usepackage{textcomp}%
\usepackage{manyfoot}%
\usepackage{booktabs}%
\usepackage{algorithm}%
\usepackage{algorithmicx}
\usepackage{algpseudocode}%
\usepackage{listings}%
\usepackage{array}
\usepackage[numbers,sort&compress]{natbib}
\usepackage[colorlinks,citecolor=blue,urlcolor=blue,linkcolor=blue]{hyperref}
\usepackage{etoolbox}
\usepackage{environ}

\smartqed  % flush right qed marks, e.g. at end of proof
\journalname{Eur. Phys. J. C}
\begin{document}
\onecolumn % Force to full-width page tracking

%%% DRAFT 28.05.2026 %%
\title{Hidden-sectors search and probe of discrete symmetries at the REDTOP experiment}
\begingroup
\renewcommand{\and}{\par\addvspace{1em}} % Forces vertical distribution if horizontal space runs out
\makeatletter
\g@addto@macro\maketitle{\allowbreak}
\makeatother
\authorrunning{REDTOP Collaboration}
\titlerunning{REDTOP Collaboration}
\author{%
  C.~Gatto\thanksref{e1,inst1,inst2} \and
  B.~Kubis\thanksref{e2,inst3} \and
  A.~Mazzacane\thanksref{e3,inst22} \and 
  M.~Zieli\'{n}ski\thanksref{e4,inst4} \and
  W.~Abdallah\thanksref{inst5} \and
  A.~Alqahtani\thanksref{inst6} \and
  D.~S.~M.~Alves\thanksref{inst7} \and
  W.~Baldini\thanksref{inst8} \and
  S.~Barbi\thanksref{inst9} \and
  B.~Batell\thanksref{inst10}
% ...
}

\thankstext{e1}{Corresponding author: gatto@fnal.gov}
\thankstext{e2}{Corresponding author: kubis@hiskp.uni-bonn.de}
\thankstext{e3}{Corresponding author: mazzacan@fnal.gov}
\thankstext{e4}{Corresponding author: marcin.zielinski@uj.edu.pl}

\institute{%
  Istituto Nazionale di Fisica Nucleare, Sezione di Napoli, 80126 Napoli, Italy\label{inst1} \and
  Northern Illinois University, Northern Illinois Center for Accelerator and Detector Development, DeKalb, IL 60115, USA\label{inst2} \and
  Universität Bonn, Helmholtz-Institut für Strahlen- und Kernphysik and Bethe Center for Theoretical Physics, 53115 Bonn, Germany\label{inst3} \and
  Jagiellonian University, Marian Smoluchowski Institute of Physics, 30-348 Kraków, Poland\label{inst4} \and
  Cairo University, Faculty of Science, Department of Mathematics, Giza 12613, Egypt\label{inst5} \and
  Georgetown University, 3700 O Street, N.W., Washington, DC 20057, USA\label{inst6} \and
  Los Alamos National Laboratory, Los Alamos, NM 87545, USA\label{inst7} \and
  Istituto Nazionale di Fisica Nucleare, Sezione di Ferrara, 44122 Ferrara, Italy\label{inst8} \and
  Università di Modena e Reggio Emilia, Modena, Italy\label{inst9} \and
  University of Pittsburgh, Pittsburgh, PA 15260, USA\label{inst10}
% ...
\and
  Fermi National Accelerator Laboratory, Batavia, IL 60510, USA\label{inst22}
% ...
}
\endgroup
%%%%%%
\date{}
\maketitle

\begin{abstract}
The $\eta$ and $\eta^{\prime}$ mesons are nearly unique in the particle universe since they are nearly Goldstone bosons, and their decay dynamics are strongly constrained. While earlier experiments collected samples of order $\sim 10^{9}$ $\eta$, the proposed REDTOP (Rare Eta Decays To Observe Physics Beyond the Standard Model) facility targets $\mathcal{O}(10^{14})$ $\eta$ and $\mathcal{O}(10^{12})$ $\eta'$, enabling broad searches for physics beyond the Standard Model. 
In this work, we present studies evaluating REDTOP sensitivity to processes that couple the Standard Model to New Physics through four portals: the Vector (dark photon), the Scalar (Higgs-mixing), the Axion-like, and the Heavy Lepton. In parallel, the proposed statistics allow precise tests of $CP$ and $T$ invariance and lepton universality and improve determinations of the $\eta/\eta'$ transition form factors, which are crucial inputs to the hadronic light-by-light contribution to the muon anomalous magnetic moment $(g-2)_\mu$. 
\end{abstract}

\tableofcontents

\clearpage % Forces the huge front matter block to flush completely
\twocolumn % Returns the rest of the paper to the journal's standard layout
\section{Introduction}\label{sec:Intro}

\begin{sloppypar}
The Standard Model (SM) is not a complete description of nature. Key open questions, such as the nature of dark energy, dark matter ~\cite{Planck:2018vyg}, the origin of neutrino mass~\cite{Gonzalez-Garcia:2002bkq}, and the baryon asymmetry of the universe~\cite{Canetti:2012zc}, point to the need for physics beyond the Standard Model (BSM). These unresolved issues suggest the existence of new particles and/or forces, which may also involve violations of discrete symmetries in nature. For instance, the observed baryon asymmetry requires violations of $C$, $CP$, and baryon number in a system that departs from thermal equilibrium~\cite{Sakharov:1967dj}. As experimental precision has improved, discrepancies between theory and measurement have become more apparent. 
Nevertheless, existing data do not clearly indicate where New Physics may lie, and the allowed mass range for new particles spans over forty orders of magnitude, making experimental discovery extremely challenging~\cite{Bertone:2004pz,Bertone:2018krk}.
However, the simple assumption that new particles reach thermal equilibrium with standard matter narrows the relevant energy range for searches to approximately keV to 100 TeV, the so-called Cold Dark Matter scenario~\cite{Tuominen:2021symmetry}.
Below this range, cosmological observations place stringent constraints on New Physics. Above $\approx$ 100 TeV, unitarity bounds limit the possible interaction strengths of such particles~\cite{Viel:2013apy,Griest:1990kh}.
As a result, the keV--100 TeV window has become the primary focus of a wide array of experiments, including both accelerator-based efforts and cosmological or astrophysical observations. Within the Cold Dark Matter framework, Big Bang Nucleosynthesis imposes strong constraints on New Physics below the MeV scale~\cite{Yeh_2022}, while searches above the GeV scale are increasingly disfavored by direct detection~\cite{XENON:2018voc} and accelerator-based experiments, particularly those at the LHC~\cite{ATLAS:2021kxv,CMS:2021ctt}. 
This leaves the MeV--GeV mass range, often referred to as the light cold dark matter regime, among the least constrained. Recent theoretical developments provide strong motivation for a more detailed exploration of this parameter space~\cite{Boehm:2003hm,Pospelov:2007mp,Knapen:2017xzo,Alexander:2016aln,Beacham:2019nyx}. 
\end{sloppypar}

A defining feature of many recent models is that interactions between New Physics and the SM occur via extremely small couplings, often on the order of $10^{-8}$ or lower~\cite{Batell:2009di,Alexander:2016aln,Essig:2013lka}. In this context, fixed-target experiments with intense beams can play a pivotal role~\cite{Proceedings:2012ulb,Batell:2009di,Essig:2013lka}. These experiments can achieve luminosities on the order of $10^{45}$ cm$^{-2}$, far beyond what is currently feasible at high-energy colliders. 
When accounting for various production mechanisms, fixed-target experiments can achieve GeV-scale production rates of neutral states that are several orders of magnitude higher than those at colliders. 
This makes them uniquely suited to explore rare processes and significantly increases the likelihood of discovering New Physics in the most elusive regions of parameter space.

\begin{sloppypar}
It is well established that light dark matter must be neutral under SM charges; otherwise, it would likely have been detected already~\cite{Goldberg:1983nd,Ellis:1983ew,Bertone:2004pz,Feng:2010gw}. Among the known particles, only the $\eta$ and $\eta'$ mesons, the Higgs boson, and the vacuum itself have all zero quantum numbers, apart from the negative parity of the pseudoscalars, a combination of properties that is exceptionally rare in nature~\cite{ParticleDataGroup:2024cfk,Feldmann:1999uf}.\footnote{In principle, also the heavy quarkonia $\eta_c$ and $\eta_b$ share these properties; due to the completely different production and decay processes, we however do not discuss these in this article.} 
The decays of the $\eta$ and $\eta'$ are flavor-conserving and proceed through neutral currents~\cite{Gasser:1984gg,Escribano:2013kba}, as charged processes are forbidden by symmetry constraints. 
Consequently, many decay modes are suppressed or forbidden at leading order in the SM, thereby enhancing their sensitivity to potential couplings to light dark-sector states~\cite{Gan:2020aco}. 
In this respect, an $\eta$/$\eta'$ factory provides a complementary environment to heavy-flavor or high-energy experiments, uniquely suited for probing rare and symmetry-violating processes.
\end{sloppypar}

An experimental program capable of collecting on the order of $10^{14}$ $\eta$ and $10^{12}$ $\eta'$ mesons would have the statistical power to test the predictions of a wide range of recent theoretical models. Such a program would enable a comprehensive exploration of the mostly unconstrained parameter space associated with all four portals connecting the Dark Sector to the SM. The proposed REDTOP experiment is designed to explore this frontier by focusing on both the search for new particles and the study of fundamental symmetry violations. With its enhanced sensitivity, REDTOP aims to improve the limits on key conservation laws by several orders of magnitude compared to previous experiments. This would extend the reach of searches for physics BSM, including dark matter, dark energy, and new fundamental forces.

\begin{sloppypar}
The REDTOP measurements will focus primarily on rare decays of the $\eta$ and $\eta'$ mesons produced by a proton or pion beam with an energy of a few GeV.  
Early measurements will focus on testing the conservation laws of $C$, $CP$, and $T$ with sensitivities several orders of magnitude beyond current experimental limits, while also searching for new particles and Light Cold Dark Matter. We have compiled a comprehensive list of potential measurements that can be performed with REDTOP, organized into six distinct categories (Table~\ref{table:decay_list}). 
While the experiment encompasses a wide range of physical goals, we highlight a few particularly promising processes.

\begin{table*}[t]
\centering
\renewcommand{\arraystretch}{1.2}
\begin{tabular}{c}
\toprule 
\textbf{\boldmath $C$-, $T$-, $CP$-violation tests}\\
\midrule
$CP$ violation via Dalitz plot mirror asymmetry: $\eta\to\pi^{+}\pi^{-}\pi^{0}$ \\
$CP$ violation (Type I \textendash{} $P$ and $T$ odd, $C$ even): $\eta\to4\pi^{0}$ \\
$CP$ violation (Type II - $C$ and $T$ odd, $P$ even): $\eta\to\pi^{+}\pi^{-}\pi^{0}$
and $\eta\to\gamma\gamma\gamma$ \\
Test of $CP$ invariance via $\mu$ longitudinal polarization: $\eta\to\mu^{+}\mu^{-}$\\
Test of $CP$ invariance via $\gamma^{*}$ polarization studies: $\eta\to\pi^{+}\pi^{-}e^{+}e^{-}$
 and  $\eta\to\pi^{+}\pi^{-}\mu^{+}\mu^{-}$ \\
Test of $CP$ invariance in angular correlation studies: $\eta\to\mu^{+}\mu^{-}e^{+}e^{-}$\\
Test of $T$ invariance via $\mu$ transverse polarization: $\eta\to\pi^{0}\mu^{+}\mu^{-}$
and $\eta\to\gamma\mu^{+}\mu^{-}$ \\
$CPT$ violation: $\mu$ polariz. in $\eta\to\pi^{+}\mu^{-}\nu$ vs $\eta\to\pi^{-}\mu^{+}\nu$
and $\gamma$ polarization in $\eta\to\gamma\gamma$\\\addlinespace
\midrule 
\textbf{Searches for new particles and forces }\\
\midrule
Scalar meson searches (charged channel): $\eta\to\pi^{0}H$, with
$H\to e^{+}e^{-},$ $\mu^{+}\mu^{-}$ \\
Dark photon searches: $\eta\to\gamma A'$ with $A'\to l^{+}l^{-}$\\
Protophobic fifth force searches : $\eta\to\gamma a$ with $a\to e^{+}e^{-}$\\
New leptophobic baryonic force searches :  $\eta\to\gamma B$, with
$B\to e^{+}e^{-},$ $\pi^{0}\gamma$\\
Indirect searches for  new gauge bosons and leptoquark:
$\eta\to\mu^{+}\mu^{-}$ and $\eta\to e^{+}e^{-}$\\
Search for true muonium: $\eta\to\gamma$ $(\mu^{+}\mu^{-})$$_{2M_{\mu}}$\textrightarrow{}
$\eta\to\gamma$ $e^{+}e^{-}$ \\
All invisible or semi-visible  $\eta$ decays
\\\addlinespace
\midrule 
\textbf{Other discrete symmetry violations} \\
\midrule
Lepton flavor violation: $\eta\to\mu^{+}e^{-} + \text{c.c.}$ \\
Double lepton flavor violation: $\eta\to\mu^{+}\mu^{+}e^{-}e^{-} + \text{c.c.}$ \\\addlinespace
\midrule 
\textbf{Other precision physics measurements }\\
\midrule
Proton radius anomaly: $\eta\to\gamma\mu^{+}\mu^{-}$ vs.\ $\eta\to\gamma e^{+}e^{-}$ \\
All unseen leptonic decay modes of $\eta/\eta'$ (SM predicts $10^{-6}-10^{-9}$) \\\addlinespace
\midrule 
\textbf{Non-$\eta/\eta'$ based BSM physics }\\
\midrule
Dark photon and ALP searches in Drell--Yan processes: $q\bar{q}\rightarrow A'/a\rightarrow l^{+}l^{-}$ \\
ALP searches in Primakoff processes: $pZ\rightarrow pZa\rightarrow l^{+}l^{-}$\\
Charged pion and kaon decays: $\pi^{+}\rightarrow\mu^{+}\nu A'\rightarrow\mu^{-}\nu l^{+}l^{-}$
and $K^{+}\rightarrow\mu^{+}\nu A'\rightarrow\mu^{-}\nu l^{+}l^{-}$ \\
Neutral pion decay: $\pi^{0}\rightarrow\gamma A'\rightarrow\gamma e^{+}e^{-}$\\\addlinespace
\midrule 
\textbf{High-precision studies on nonperturbative QCD physics }\\
\midrule
Chiral perturbation theory \\
Nonperturbative QCD \\
Isospin breaking due to the $u$-$d$ quark mass difference \\
Octet--singlet mixing angle \\
Electromagnetic transition form factors (important input for $g-2$) \\\addlinespace
\bottomrule 
\end{tabular} 
\caption{Overview of the physics program of REDTOP, listing the main research topics and the decay channels accessible for their investigation.}
\label{table:decay_list} 
\end{table*}

A particularly compelling class of theories beyond the SM involves hidden sectors, where new, weakly coupled degrees of freedom remain inaccessible to conventional high-energy experiments. These models often postulate the existence of so-called portals~\cite{Batell:2009di}, which mediate interactions between the visible and hidden sectors via renormalizable operators. 
For clarity, the vector portal allows mixing between a new $U(1)$ gauge boson, or dark photon, and the photon. The scalar and pseudoscalar portals involve couplings of light Higgs-like or Axion-Like Particles to SM fields, and the neutrino portal introduces heavy neutral leptons that mix with SM neutrinos~\cite{Patt:2006fw,Holdom:1985ag,Arkani-Hamed:2008hhe,Jaeckel:2010ni,Bauer:2017ris,Minkowski:1977sc,Boyarsky:2009ix}.
The REDTOP experiment aims to push current branching-fraction limits below  $10^{-9}$, reaching below $10^{-11}$ for most $\eta$ decay modes. 

Among the many processes accessible at REDTOP (as shown in Table~\ref{table:decay_list}), a few stand out for their particularly high discovery potential:
\begin{itemize}

\item{$\eta \to \pi^+ \pi^- \pi^0$: {\textit{CP Violation in the Dalitz Plot}}}.
The observation of the Dalitz plot asymmetry would be direct evidence of $C$ and/or $CP$ violation~\cite{Lee:1965zza,KLOE-2:2016zfv,Gardner:2019nid,Akdag:2021efj,Akdag:2022sbn,Shi:2024yfa}.

\item{$\eta \to \mu^+ \mu^- e^+ e^-$: \textit{CP Violation in Angular Asymmetry}}.
This decay is sensitive to $CP$-violating angular correlations between the $e^+e^-$ and $\mu^+\mu^-$ decay planes~\cite{Zillinger:2022eva,Escribano:2022zgm,BESIII:2013tjj,BESIII:2020elh}. 

\item {$\eta \to \mu^+ \mu^- X$:
\textit {CP Violation via Transverse or Longitudinal Muon Polarization}}.
In this class of decays, the muon pair can be produced either in a $^1S_0$ state (CP-conserving) or a $^3P_0$ state (CP-violating)~\cite{Geng:1990dw}. 
A nonzero measurement of muon polarization is a straightforward indication of $CP$-violation originating from BSM processes~\cite{Sanchez-Puertas:2018tnp}.

\item{$\eta \to \gamma A'$, with $A' \to e^+ e^-$ or $\mu^+ \mu^-$: \textit{Dark Photon and Light Gauge Boson Search}}. 
This decay allows for searches for dark photons ($A'$) or light vector bosons, such as a potential mediator of a fifth, multi-weak force~\cite{Fayet:1990wx,Fayet:2007ua,Feng:2008ya,Reece:2009un,HADES:2013nab,WASA-at-COSY:2013zom,KLOE-2:2011hhj}.

\item{$\eta \to \pi^0 H$, with $H \to e^+ e^-$: \textit{Higgs-Like Scalar Search}}.
This channel probes light scalar bosons that couple to SM particles via mixing with the Higgs sector. 
Such scalars appear in extended Higgs models, dark scalar portals, and light dark matter mediator scenarios. The decay $\eta \to \pi^0 H$ provides sensitivity to scalar masses below $\sim 500$~MeV, complementing searches in kaon and $B$-meson decays~\cite{Dobrich:2018jyi}. 

\item{\textit{Axion-Like Particle (ALPs) Searches}}.
Search for axion-like particles (ALPs), including the QCD axion, as signatures of BSM physics~\cite{Bauer:2017ris}. 
They can be produced in $\eta^{(\prime)} \to M_1 a$ or $\eta^{(\prime)} \to M_1 M_2 a$ decays (with $M_i$ representing mesons) providing model-independent bounds on the Peccei-Quinn scale ($f_a$)~\cite{Peccei:1977hh,Peccei:1977ur,Weinberg:1977ma,Wilczek:1977pj}. Specific channels include $\eta \to \pi^0 \pi^0 a$ or $\eta' \to \pi^+ \pi^- a$ decays, where $a$ is an ALP that decays into $e^+ e^-$~\cite{Aloni:2018vki,Alves:2017avw,Alves:2020xhf,Alves:2024dpa}. For other portals, the ALP can decay to $\gamma\gamma$, $3\pi$, or $\pi\pi\eta$ without an initial $\eta$ or $\eta'$~\cite{Batell:2009di}.

\end{itemize}

The broad physics program outlined above, ranging from searches for light vector and scalar mediators, tests of $CP$ violation in multilepton final states, precision studies of $\eta\to3\pi$, and the exploration of ALPs signatures, places stringent and complementary requirements on the experimental apparatus.  
These channels typically involve low-momentum leptons and photons, rare electromagnetic transitions, and, in several cases, the possibility of long-lived particles or detached decay vertices. Example searches for light vector particles require excellent sensitivity to very low-energy $e^{+}e^{-}$ pairs with small opening angles. Similarly, models predicting a piophobic QCD axion with a mass of up to a few tens of MeV call for a calorimeter capable of reconstructing extremely low-mass diphoton resonances and low-momentum leptons.  
Other BSM scenarios allow for long-lived states whose decays would manifest as displaced vertices, since no SM process at REDTOP energies is expected to produce such signatures. Even a single well-reconstructed detached vertex would be indicative of physics beyond the SM. 
This places stringent requirements on the vertex detector, which must achieve spatial resolutions at the level of a few tens of microns. 
Similarly, invisible or semi-visible $\eta$ decays are strongly suppressed in the SM. They can be very constraining for dark-matter models~\cite{Ema:2020ulo}.
Finally, the experiment must operate in an exceptionally high-rate environment. The inelastic interaction rate of the proton beam on the target is expected to approach 700 MHz. Although the average event multiplicity is modest, the sheer rate demands a sophisticated trigger and data-acquisition strategy capable of suppressing overwhelming hadronic background while retaining efficiency for rare processes. As a consequence, the REDTOP detector must combine excellent vertexing, a low material budget, efficient lepton and photon identification, and the capability to operate in a high-rate hadronic environment at GHz rates.
These considerations define the guiding principles behind the detector design presented in the next section.
\end{sloppypar}

In this paper, we present a comprehensive assessment of the expected physics performance of the REDTOP experiment. 
Motivated by the physics outlined above, we begin with a discussion of the detector requirements and the conceptual design of the proposed apparatus, highlighting the functionality of its principal subsystems and their roles in reconstructing rare $\eta$ and $\eta^{\prime}$ decays. Building on these design considerations, we present detailed simulation studies covering several representative physics scenarios that span the main categories of processes discussed in the introduction. These studies, which incorporate realistic assumptions about detector geometry, material budget, reconstruction performance, and background conditions, are used to evaluate the achievable sensitivities for each decay channel. The subsequent section describes the expected beam configurations and intensity requirements needed to fully exploit the experimental physics potential, together with the constraints imposed by the high-rate environment of a fixed-target facility. Finally, the paper concludes with a roadmap for REDTOP and an outlook on the steps toward realizing the full physics program.

\section{REDTOP Detector Concept and Design}\label{theDetector}
\begin{sloppypar}

The physics objectives of REDTOP, as outlined in the previous section, necessitate a detector capable of reconstructing rare $\eta$ and $\eta^{\prime}$ decays with high precision in a high-rate fixed-target environment~\cite{Gan:2020aco}. Given the quantum numbers of the $\eta$ and $\eta^{\prime}$ mesons, hadroproduction represents the only viable mechanism to achieve the event yields required by the REDTOP physics program~\cite{Gatto:2019dhj}. To address these challenges, the experiment adopts a low-mass, high-granularity design optimized for efficient tracking, calorimetry, and particle identification. Dedicated target and vertexing systems are combined with a central tracking volume located within a solenoidal magnetic field, surrounded by fast timing detectors, particle identification systems, and high-performance electromagnetic and hadronic calorimetry. 
\end{sloppypar}

\begin{sloppypar}
The inelastic interaction rate between the beam and the target is expected to approach the GHz scale, which calls for detectors with excellent time resolution. 
For this purpose, a multi-level trigger and data acquisition architecture is designed to cope with large event rates and efficiently select rare decay signatures. 
\begin{figure*}[t]
  \centering
  \includegraphics[width=0.9\textwidth]{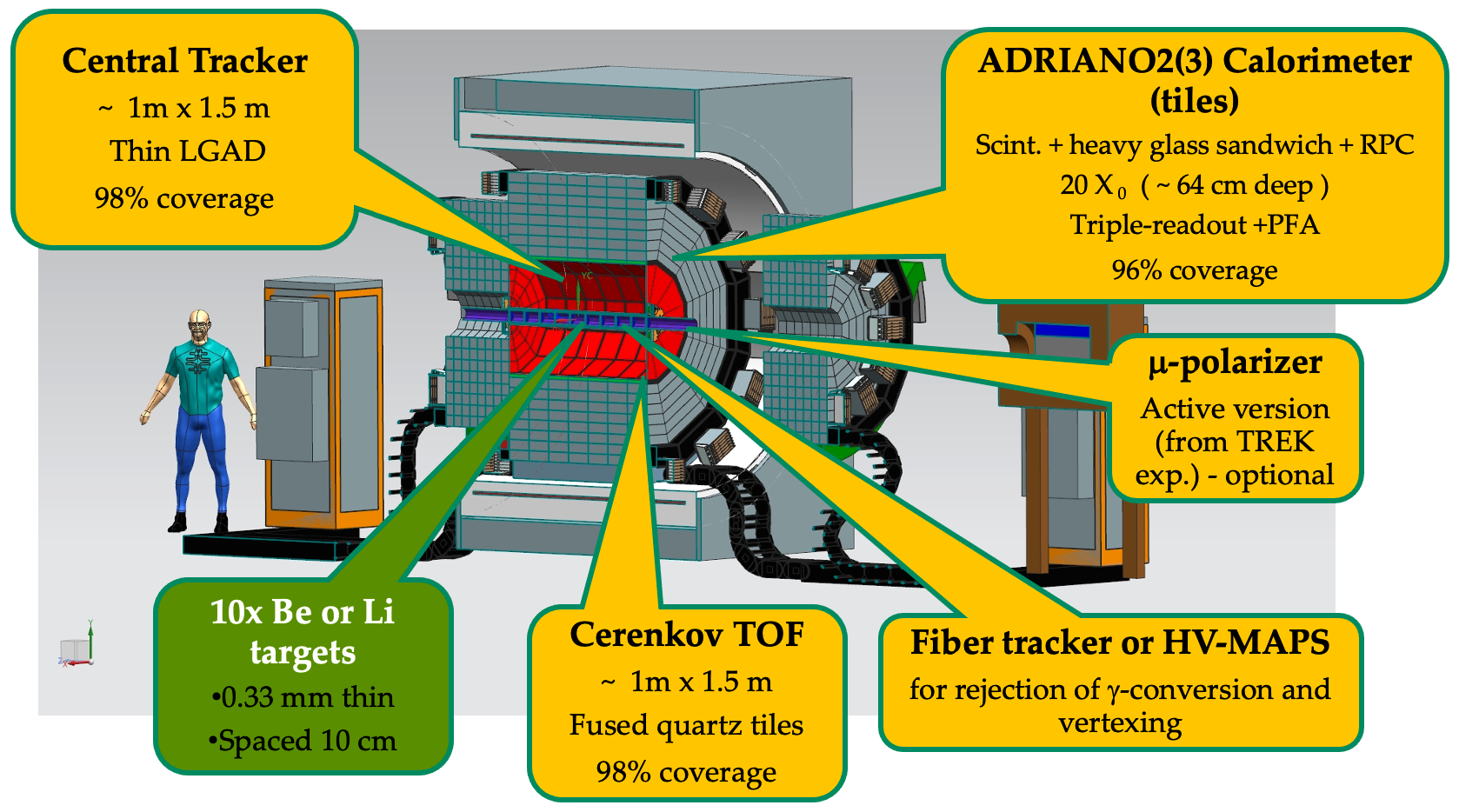}
  \caption{Baseline schematic layout of the REDTOP detector.  The proton beam interacts with a segmented target of ten thin (0.33~mm) Be or Li foils, separated by 10~cm, producing $\eta$ and $\eta'$ mesons. Charged-particle tracking is provided by a thin LGAD central tracker (CTOF) ($\sim$1~m $\times$ 1.5~m, 98\% coverage), complemented by an inner fiber tracker (or HV-MAPS) for the rejection of photon conversions and  secondary-vertex reconstruction. Particle identification relies on a fused-quartz Čerenkov Time-of-Flight system (98\% coverage), while calorimetry is performed by the ADRIANO-2(3) dual-readout calorimeter (scintillator, heavy glass, RPC, 96\% coverage) with triple-readout particle-flow analysis. An optional muon polarimeter (based on the TREK design) extends the physics reach to $CP$-violation studies in $\eta \to X \mu^+ \mu^-$ decays. This configuration is the reference geometry for all simulation studies presented in the following sections.}
  \label{fig:reddetector}
\end{figure*}
At the same time, the hadronic production environment is dominated by a very large flux of slow baryons, primarily protons and neutrons, which are not of direct physics interest. This requires a tracking system designed to minimize its impact on occupancy and background. Finally, the signatures of many targeted rare processes and possible new physics scenarios involve leptons and photons in the final state, making high-performance 
calorimetry and robust particle identification essential components of the experiment. A general schematic of the detector concept is shown in Fig.~\ref{fig:reddetector}. This section provides an overview of the concept of the REDTOP detector and the functionality of its main subsystems.
\end{sloppypar}

\subsection{Target Systems}

The choice of the REDTOP target system is closely connected to the beam configuration discussed in Sec.~\ref{sec:beam}. Two beam scenarios are currently under consideration, based on either proton or pion beams, and the target design has been developed accordingly. The baseline options include a system of multiple thin, solid foils of lithium or beryllium, as well as a liquid deuterium (LDe) target.

The solid-target configuration consists of a set of thin foils, typically  5-10 lithium or beryllium disks, each 770~$\mu$m (240~$\mu$m for Be) and about 1~cm in diameter, aligned along the beam axis and suspended inside a low-mass beryllium or carbon fiber beam pipe. This layout minimizes the material traversed by the decay products and enables precise reconstruction of primary interaction vertices~\cite{LHCb:2008vvz,HERA-B:2000smc}. In this configuration, the inelastic interaction probability per incident proton is on the order of a few percent, while only a small fraction of the beam is absorbed, resulting in a negligible thermal load per foil. The separation of multiple thin targets along the beam direction further improves vertex discrimination and background suppression compared to extended or gaseous targets. Lithium foils offer a modest reduction in neutron background, producing, on average, about two neutrons per event compared to approximately three in the case of beryllium~\cite{GEANT4:2002zbu}. Although this difference has only a minor impact on the overall sensitivity to BSM searches, the higher chemical reactivity of lithium introduces additional handling and R\&D challenges.

\begin{sloppypar}
An alternative solution based on an LDe target has also been investigated. The LDe target is enclosed in a Kapton cylinder approximately 5~cm long with a wall thickness of 50~$\mu$m. For a beam intensity of $10^{11}$ protons on target per second, the expected interaction rate reaches about 420~MHz, with roughly 1\% of the interactions occurring in the Kapton windows. The main advantage of this option is the significantly lower hadronic background. Simulations performed with \emph{GenieHad-UrQMD}~\cite{Bass:1998ca,Bleicher:1999xi} indicate that, at 1.8~GeV and in events without $\eta$ production, the LDe target produces, on average, about 2.0 charged particles (mainly protons and pions) and 1.4 neutral particles per event, compared to approximately 3.8 charged and 3.0 neutral particles for a beryllium target under the same conditions.
An additional advantage of an LDe target is that it can be used with a pion beam, where available, to generate tagged or semi-tagged $\eta$ and $\eta^{\prime}$ mesons. 
A tagged $\eta/\eta^{\prime}$ factory has the ability to probe invisible and semi-visible decays, thereby broadening the search for BSM physics. 
Furthermore, the hadronic background generated by a pion beam is considerably lower than that from a proton beam~\cite{CrystalBall:2004qln}.
The liquid-deuterium option, however, introduces additional mechanical and operational complexity and increases the multiple scattering experienced by the $\eta$ and $\eta^{\prime}$ decay products as they traverse the target volume. Similar cryogenic systems have nevertheless been successfully operated in previous experiments~\cite{Mu3e:2020gyw}.
\end{sloppypar}

\subsection{The Vertex Detector}
The Vertex detector plays a crucial role in the tracking and background rejection strategy of REDTOP. The main operating purpose of the vertex detector is to deliver precise spatial information that is essential for disentangling the primary interaction region from the secondary decay detached vertices in a high-rate hadronic environment~\cite{H1:1996jzy,Aaij:2014zzy}. 
In particular, it enables the identification of events with detached secondary vertices, 
contributing to the reconstruction of charged tracks originating from the target and rejecting photons that are converted in the target material. Its proximity to the interaction point also allows the reconstruction of tracks with very low transverse momentum.
The most critical design challenges are the material budget and the detector proximity to the interaction point. Minimizing the material budget is essential to reduce multiple scattering, which is the dominant source of resolution degradation in the REDTOP momentum range. Excess material would also increase the background from photon conversions.

These conditions impose technically demanding requirements on the vertex detector. 
In particular, the detector must provide a spatial resolution near the interaction point better than 20 $\mu$m while maintaining a low material budget of the order of 0.1\% $X_{0}$ per layer. 
Precise timing resolution on the order of nanoseconds is required to cope with the high interaction rate, and the sensors must withstand radiation hardness up to $\sim5\times10^{6}$ cm$^{-2}$ s$^{-1}$ (1~MeV neutron equivalent), or $\sim5\times10^{13}$cm$^{-2}$ integrated over one year. 
The detector geometry is optimized to provide solid-angle coverage exceeding 90\%, which will ensure a high efficiency of track finding. 

\begin{sloppypar}
These requirements are comparable to those of the pixel tracker in the Mu3e experiment at PSI. For reference, Mu3e operates with a muon stopping rate of $\sim$100 MHz (Phase-1) to $\sim$2000 MHz (Phase-2), while REDTOP expects an inelastic event rate of 500--700 MHz.
However, REDTOP tracks reach higher momenta (up to $\sim$1.1 GeV/$c$ for pions and $\sim$2.2 GeV/$c$ for protons), compared to a maximum of $\sim$53 MeV/$c$ in Mu3e. As a result, REDTOP adopts the same HV-MAPS technology as Mu3e, using the MuPix sensor~\cite{MU3ERP2012}. This technology has demonstrated a time resolution of 6--8~ns with proper time-walk correction~\cite{Rudzki:2021smh}, which is well-matched to REDTOP's requirements.
\begin{center}
\begin{figure}[t]
\centering{}\includegraphics[width=0.42\textwidth]{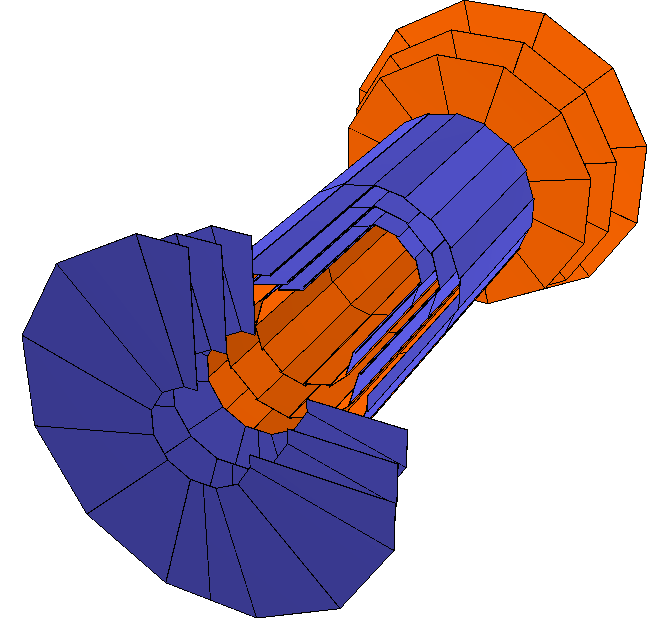}
\caption{\label{fig:ITS3}Schematics of the vertex detector of REDTOP.}
\end{figure}
\par\end{center} 

The REDTOP vertex detector consists of three layers of HV-MAPS sensors in both the barrel and endcap regions. The innermost layer has a radius of 2.4~cm from the beam axis, while the outermost layer extends to about 4.2~cm. The total detector length is 28 cm in the current layout, providing solid-angle coverage of approximately 91\%. Pixel size is optimized for the angular deflection due to multiple scattering of particles with momenta on the order of 100~MeV/$c$. This leads to a preferred pixel size in the range of 30--50  $\mu$m (layer-dependent), consistent with Mu3e specifications. Hit timestamps are derived from an internal phase-locked loop (PLL) operating at 625 MHz. With offline time-walk corrections, the MuPix sensor has achieved a timing resolution of 6 ns~\cite{Rudzki:2021smh}. A schematic of the REDTOP vertex detector layout is shown in Fig.~\ref{fig:ITS3}. The total power consumption of the vertex detector is approximately 0.8~kW, or about 400~mW/cm$^{2}$~\cite{Mu3e:2020gyw}. 
Cooling is achieved using ambient-condition cold gas, which is sufficient to maintain the detector temperature below 70°C matching the glass-transition temperature of the adhesives used in its construction.
\end{sloppypar}

\subsection{The Central Tracker}
\begin{sloppypar}
A new generation 4D tracking system is proposed for REDTOP to complement the vertex detector in reconstructing charged tracks and to assist the Central Time-of-Flight system in distinguishing leptons from hadrons. At the same time, the material budget must be kept low to minimize multiple scattering. 
The main requirements for the central tracker are:
\begin{itemize}
\item Transverse momentum resolution: $\sigma_{P_{T}}/P_{T}^{2} \sim 1 \times 10^{-2}$ GeV$^{-1}$ at $P_{T} = 1$ GeV;
\item Material budget: $\sim$0.1\% $X_{0}$ per layer;
\item Time resolution: $<$30 ps per layer.
\end{itemize}

Current Low Gain Avalanche Detector (LGAD) technology meets REDTOP performance targets in both timing and spatial resolution. The ATLAS and CMS experiments have incorporated dedicated LGAD layers in their forward regions to provide precision timing information and mitigate pileup~\cite{ATLAS-HGTD-TDR,Butler:2019rpu}. These detectors are built to withstand radiation fluences in excess of $10^{15}$n$_{\text{eq}}$/cm$^2$, and feature spatial pixel sizes of approximately $0.3 \times 0.3$ mm$^2$. For REDTOP, the pixel size is determined by the deflection of $\mathcal{O}(100~\text{MeV}/c)$ particles due to multiple scattering between layers. This leads to an optimal pixel size of 0.3--0.6~mm (layer dependent), which is slightly smaller than the CMS Endcap Timing Layer (ETL) configuration.

\begin{figure}[th]
\centering
\includegraphics[width=0.90\columnwidth]{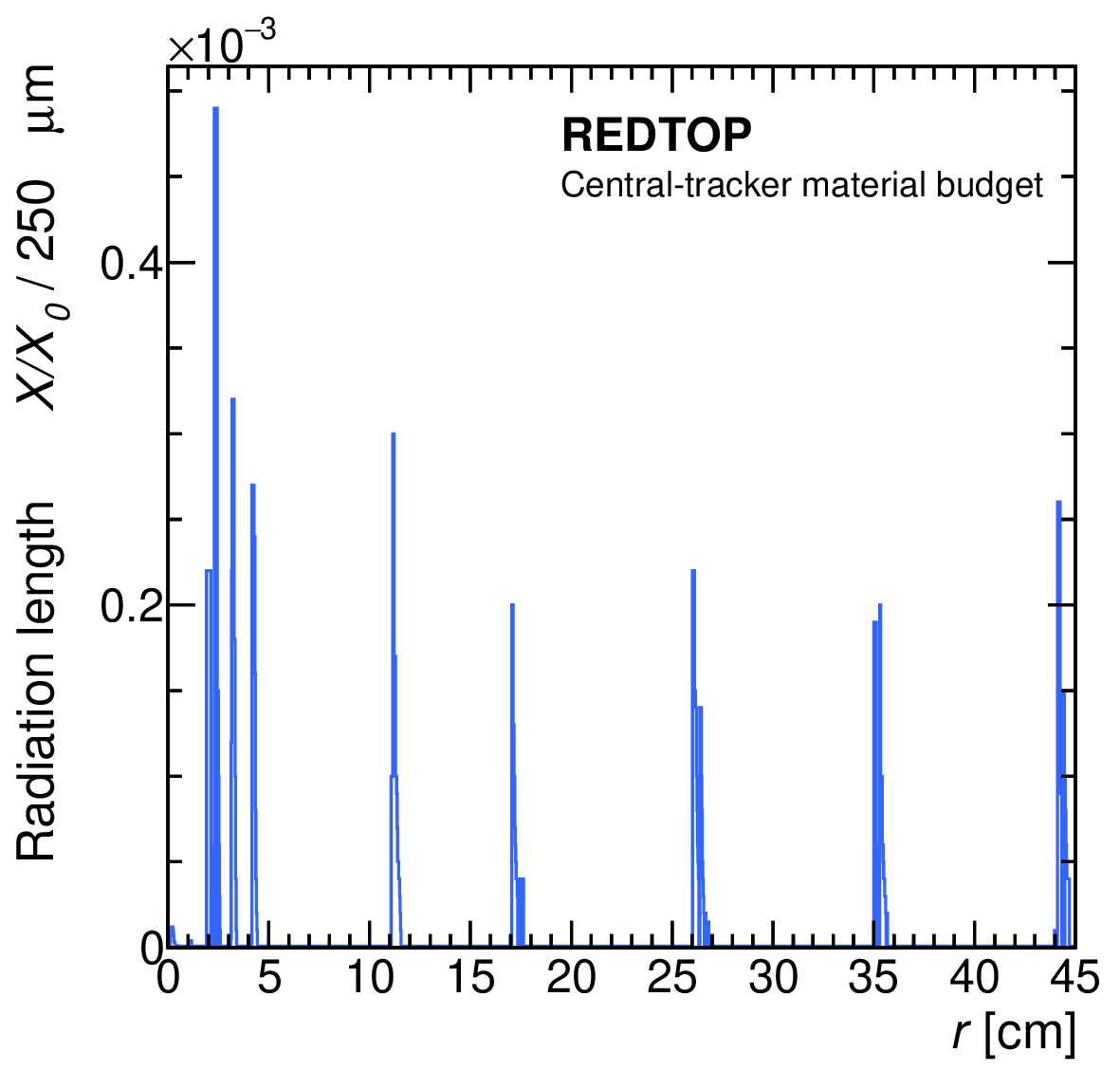}
\includegraphics[width=0.90\columnwidth]{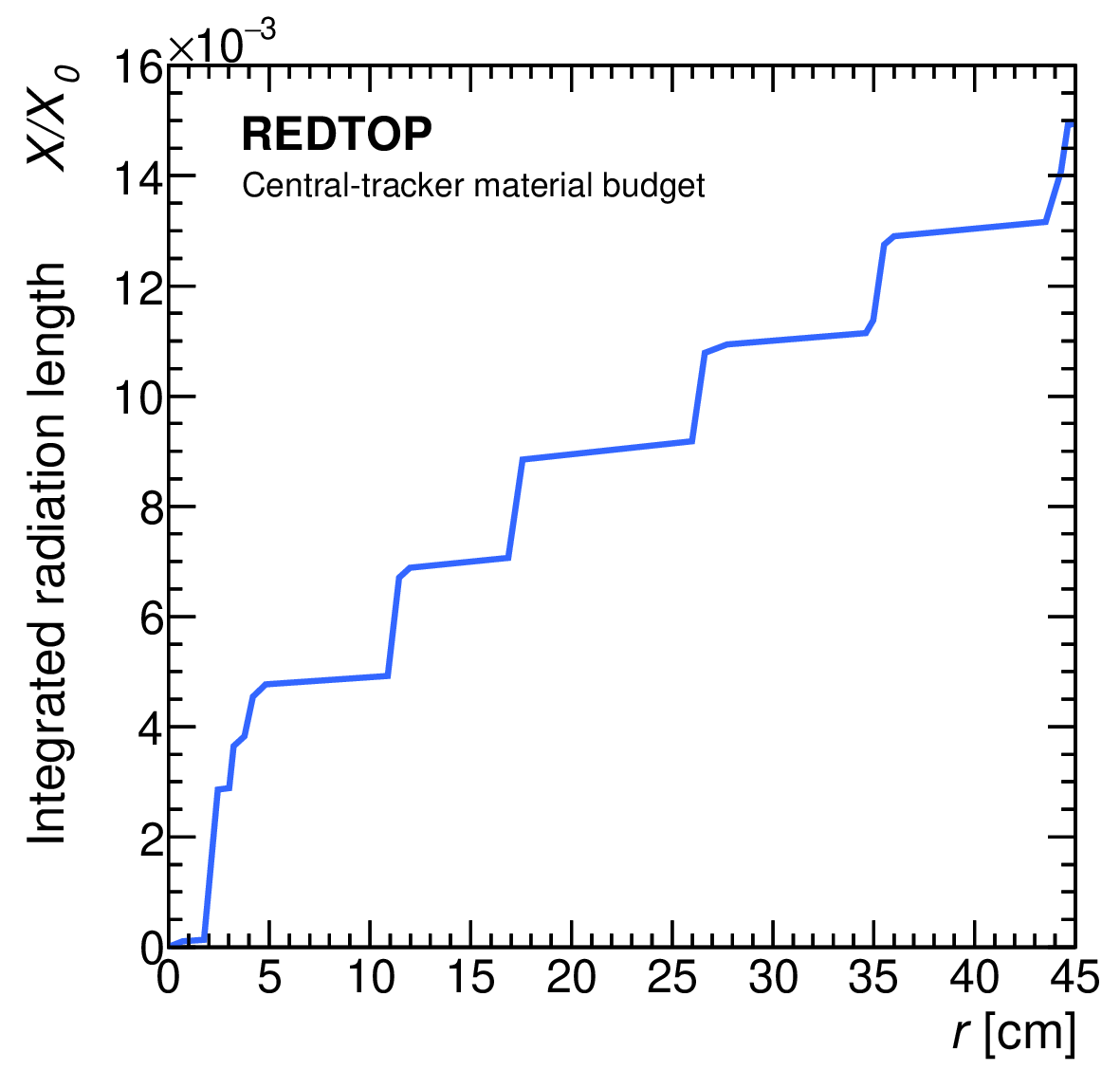}
\caption{Material budget of the  REDTOP Central Tracking system 
as a function of the transverse radius $r$ from the beam axis. 
Top panel: Differential radiation length $X/X_0$ evaluated in 
250~$\mu$m radial bins, where peaks at $r \lesssim 5$~cm correspond to 
the target and beam pipe region, while the five peaks at 
$r \approx 11$, 17, 26, 35 and 45~cm correspond to the LGAD layers 
of the Central Tracker. 
Bottom panel: Integrated material budget $X/X_0(r)$ for the same 
geometry, reaching $\sim 1.5\%~X_0$ at the outer tracker boundary.}
\label{fig:Material-budget}
\end{figure}
\begin{figure}[t]
\centering
\includegraphics[width=0.42\textwidth]{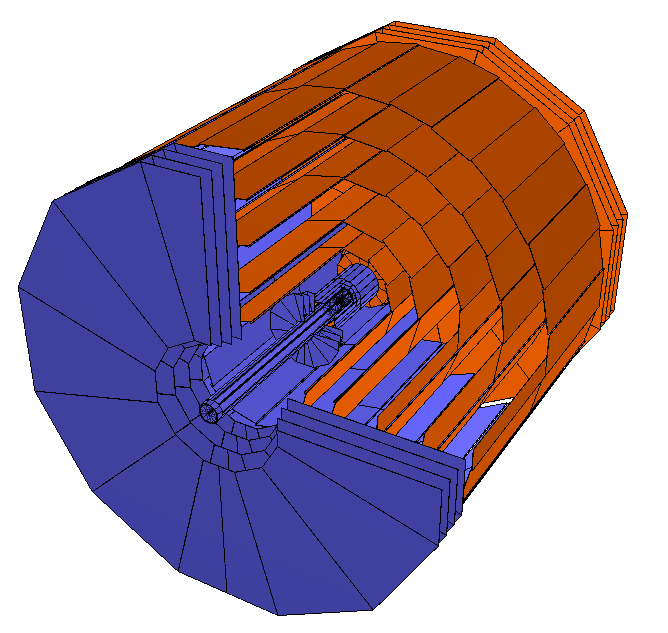}
\caption{\label{fig:LGAD}Schematics of the central tracking system of REDTOP.}
\end{figure}

The largest contribution to the material budget in the HL-LHC LGAD timing layers comes from cooling and mechanical support structures; the sensors and readout ASICs contribute only a small fraction. To stay within REDTOP’s strict material limits, only passive cooling will be employed. Carbon-fiber support structures—similar to those used in the HL-LHC pixel detector upgrades—are both lightweight and thermally conductive. Minimizing heat dissipation from the readout electronics is also a critical design requirement. AC-coupled LGAD (AC-LGAD) sensors can achieve total thicknesses as low as $\sim$100 $\mu$m (including base and sensor).
Figure~\ref{fig:Material-budget} shows an estimate of the material budget as a function of the transverse radius for the full tracking system. The beam pipe contributes approximately 0.2\% $X/X_{0}$, the vertex detector about 0.1\% $X_{0}$ per layer, and the central tracker around 0.2\% $X_{0}$ per layer. A schematic of the REDTOP central tracker layout is shown in Fig.~\ref{fig:LGAD}.

The performance of the REDTOP tracking system has been studied in detail with Monte Carlo simulations to assess the impact of the material budget and pixel size on particles with momenta ranging from about 30~MeV to a few hundred MeV. Simulated tracks were reconstructed using a Kalman filter, a kink finder, and a Billoir vertex fit. The transverse momentum resolution was found to be $\sim 1 \times 10^{-2}$~GeV$^{-1}$ at $p_{T} = 1$~GeV for all particles except electrons. For the latter, the $p_{T}$ resolution was found to be approximately a factor of two worse. The track impact parameter resolution was found to be $\sim 0.12$~mm for all particles except electrons, for which the spatial resolution was approximately four times worse. The resolution for detached vertices was found to be $\sim 0.3$~mm.
\end{sloppypar}

 \subsection{The Threshold Čerenkov Time-of-Flight (Čerenkov-TOF)}
\begin{sloppypar} 
Efficient identification of particles in the presence of a larger baryonic background is an important requirement for the REDTOP detector. This role is fulfilled by the Threshold Čerenkov Time-of-Flight system, which combines Čerenkov light detection with precise timing measurement. 
The Čerenkov-TOF system is a thin Čerenkov radiator with a relatively low refractive index, positioned between the central tracker and the calorimeter. By design, is sensitive only to particles that are above the Čerenkov threshold while rejecting slower particles (mostly protons and pions), which represent the largest contribution to the background. Table~\ref{table:TCR-C-threshold} summarizes the Čerenkov momentum thresholds for various particle species.
In addition, Čerenkov-TOF measures the Time-Of-Flight (TOF), which contributes to the early stage of event selection. In particular, Čerenkov-TOF contributes to the Level-0 trigger, helping to suppress backgrounds from  baryons, which are abundant but not of primary interest. Moreover, its low material budget allows it to act as a low density pre-shower detector, complementing the calorimeter energy measurements. 
The Čerenkov-TOF detector is design to operate with a refractive index: $n_D < 1.45$, making it insensitive to protons with kinetic energies below $400$~MeV. To cope with the high rates, a time resolution of the order of 30~ps and a detector response within 100~ns, while maintaining high granularity to reduce pile-up, are required.

The baseline Čerenkov-TOF design consists of two layers of small quartz tiles. 
Each tile has the same dimensions as the ADRIANO2 tiles: $3 \times 3 \times 1$ cm$^3$. Čerenkov light is read out by four NUV-sensitive SiPMs, actively ganged and optically coupled to each tile. The SiPMs are soldered to a PCB, which is connected via a flexible PCB (flexprint) to an ASIC on the readout board. In test beams conducted by the T1604 Collaboration, an intrinsic time resolution of approximately 50~ps was achieved for individual tiles. With two layers as proposed, the system is expected to meet the time resolution requirements of the Čerenkov-TOF. The readout and resolution specifications for the Čerenkov-TOF are similar to those of the CMS MTD Endcap Timing Layer (50 ps per hit and 35 ps per track, with readout based on a 320 MHz clock). As such, the Čerenkov-TOF could adopt the same electronics and architecture based on the ETROC2 chip, with only minor modifications. The primary difference between REDTOP Čerenkov-TOF and CMS ETL is the sensor technology (SiPMs vs. LGADs) however, this can be compensated for with an appropriate charge-matching interface between the SiPMs and the ASIC.
\begin{table}
\centering
\renewcommand{\arraystretch}{1.2}
\begin{tabular}{cc}
\toprule
PID & Momentum threshold [$\text{MeV}/c$]\\
\midrule
electron & $0.4$\\ 
muon & $100$\\ 
pion & $130$\\ 
proton & $870$\\
\bottomrule 
\end{tabular}
\caption{Momentum thresholds for \u{C}erenkov light emission in the 
fused-quartz tiles ($n \approx 1.46$) of the REDTOP Čerenkov-TOF 
detector, for selected charged-particle species. The hierarchy 
of thresholds across more than three orders of magnitude in mass 
provides efficient particle identification in the momentum range 
relevant to $\eta$ and $\eta'$ decay studies.}
\label{table:TCR-C-threshold}
\end{table}

Further improvements in particle identification could be achieved by reducing the Čerenkov-TOF timing resolution requirement to less than 30~ps. Preliminary studies by the T1604 Collaboration suggest that this level of precision may exceed the limits of the current Čerenkov-TOF technology. To address this, an optional Timing Layer ($TL$) using AC-LGAD sensors similar to those planned for the Electron-Ion Collider (EIC) could be added. 
In the proposed configuration, the $TL$ would be mounted on the Čerenkov-TOF structure, adding only a small contribution to the overall material budget, with negligible impact on the performance of the downstream ADRIANO2 calorimeter.
\end{sloppypar}

\subsection{\label{subsubsec:Calorimeter}Electromagnetic and Hadronic Calorimeter: ADRIANO2 and ADRIANO3}
\begin{sloppypar}
The calorimeter system constitutes a central component of the REDTOP detector, providing precise measurements of electromagnetic and hadronic energy deposition over a large solid angle. In addition to measuring the energy and time of arrival of incoming particles, the REDTOP calorimeter system plays a crucial role in particle identification (PID) and participates in all trigger levels. Therefore, it must provide pre-processed information on very short timescales. The primary challenge for the calorimeter is to distinguish between electromagnetic (EM) and hadronic showers, particularly those initiated by neutral particles such as photons and neutrons, which are not directly detected by other subsystems. A multiple-readout calorimetric approach effectively addresses these requirements.

Based on realistic simulations and dedicated design studies, the main  performance requirements of the calorimeter system have been established. The  electromagnetic energy resolution is required to satisfy $\sigma_{E}/E < 3\%/\sqrt{E}$, ensuring the precise reconstruction of photons and electrons over the relevant energy range.
The detector geometry must provide high granularity in order to operate efficiently at high interaction rates. A particle identification  efficiency exceeding 99\% for separation between electromagnetic  and hadronic particles is targeted. In addition, the calorimeter is required to achieve a time resolution better than 80~ps per cell for signals  corresponding to at least seven photoelectrons, with an overall detector response time below 100~ns. The shower reconstruction efficiency is expected to exceed 50\% for energies above 20~MeV and more than 90\% for $E > 80$ MeV.

Energy resolution is critical  in \textit{bump-hunting} analysis techniques, as narrow resonances are better resolved over the nonresonant background. On the other hand, high granularity improves the separation between gamma and neutron induced showers, which is essential for reducing background in BSM searches. High energy and spatial resolution are particularly important for identifying $\pi^0$ mesons decaying primarily into $\gamma\gamma$ or $\gamma e^{+}e^{-}$, as they are a major source of combinatorial background. High granularity not only enhances spatial resolution but also improves timing performance because smaller calorimeter cells reduce photon path lengths before being captured by the photo-sensors, lowering signal jitter and improving time resolution. 

At REDTOP energies, showers from photons and neutrons are nearly indistinguishable using purely topological criteria. However, multiple-readout calorimetry can effectively separate electromagnetic, hadronic, and neutron showers by comparing signals from different readout channels. For BSM final states with high-energy photons, detector optimization studies show that $\sim$89\% of barrel (endcap) electromagnetic showers deposit 80\% of their energy within the first 6.5 (11) radiation lengths ($X_0$), and 90\% within the first 7.8 (17) $X_0$. Therefore, REDTOP implements a front section dual-readout calorimeter and a rear section triple-readout calorimeter.

The calorimeter is therefore organized into two longitudinal sections, optimized for complementary functions. The inner section (ADRIANO2) is primarily dedicated to precise electromagnetic energy measurement and early shower discrimination, while the outer section (ADRIANO3) enhances hadronic containment and neutron identification capabilities. 
The innermost section, corresponding to $\sim$17 cm (28 cm) in the barrel (endcap), or $\sim$8 (13) $X_0$, employs the ADRIANO2 dual-readout technique~\cite{Gatto:2012jpa,Gatto:2015gna}. It consists of alternating tiles of:
\begin{itemize}
    \item SF57 lead-glass ($1\times3\times3~\text{cm}^3$), coated with diffuse ($\text{BaSO}_4$) or reflective metallic layers (e.g., Ag, Al, Mo, W) to optimize Čerenkov light collection,
    \item scintillating plastic tiles ($0.4\times3\times3~\text{cm}^3$), individually read out by on-tile SiPMs.
\end{itemize}
This dual-readout structure enables precise electromagnetic and hadronic separation, as well as rapid timing information.
While the outer section has $\sim$53 cm (67 cm) deep in the barrel (endcap), it implements the ADRIANO3 triple-readout concept. It includes:
\begin{itemize}
    \item alternating layers of ADRIANO2 tiles using 1.5 cm thick ZF2 glass in place of SF57,
    \item a thin Gd-doped glass RPC layer (added every four tile layers),
    \item 5 mm steel passive absorber plates.
\end{itemize}
This structure increases the calorimeters interaction length, improving hadronic containment. The combined depth of the two sections corresponds to approximately 25 (35) $X_0$ and 2.8 (4.0) nuclear interaction lengths ($\lambda_I$) in the barrel (endcap).

\begin{figure}[t]
\centering
\includegraphics[width=0.43\textwidth, height=0.18\textheight]{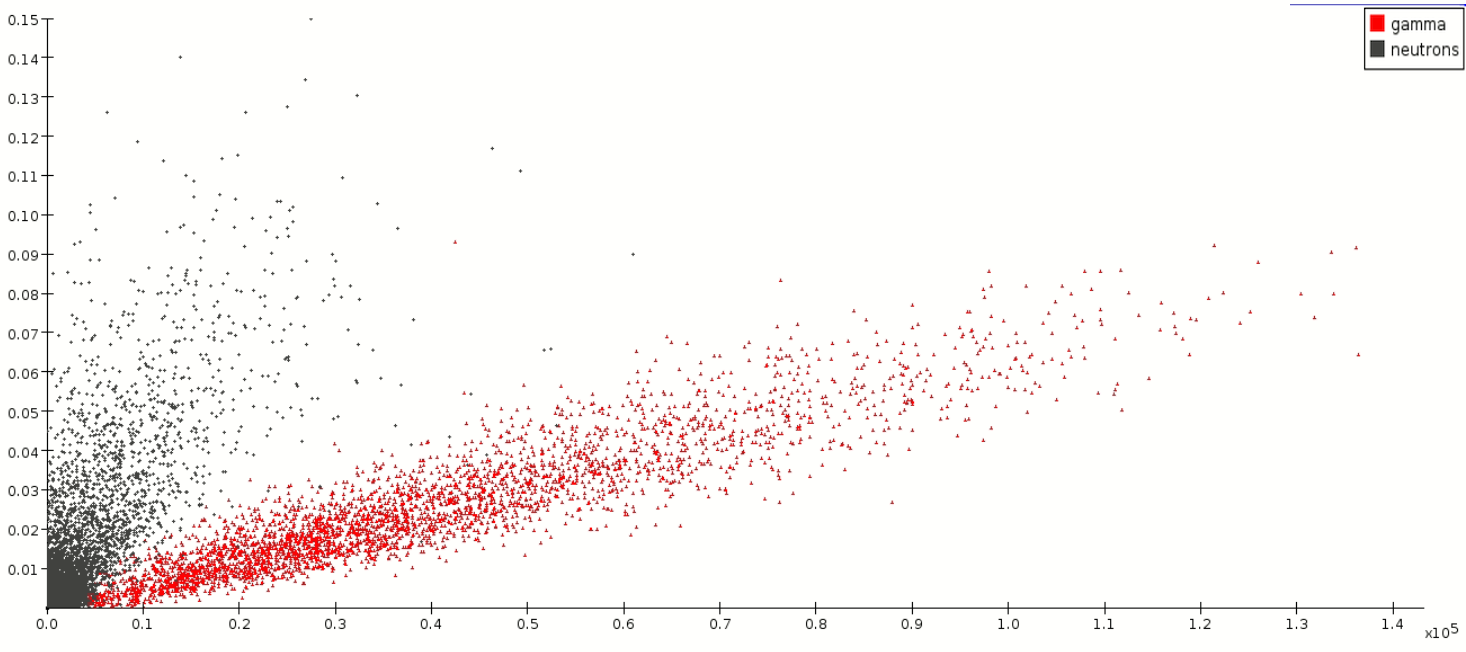}
\includegraphics[width=0.43\textwidth]{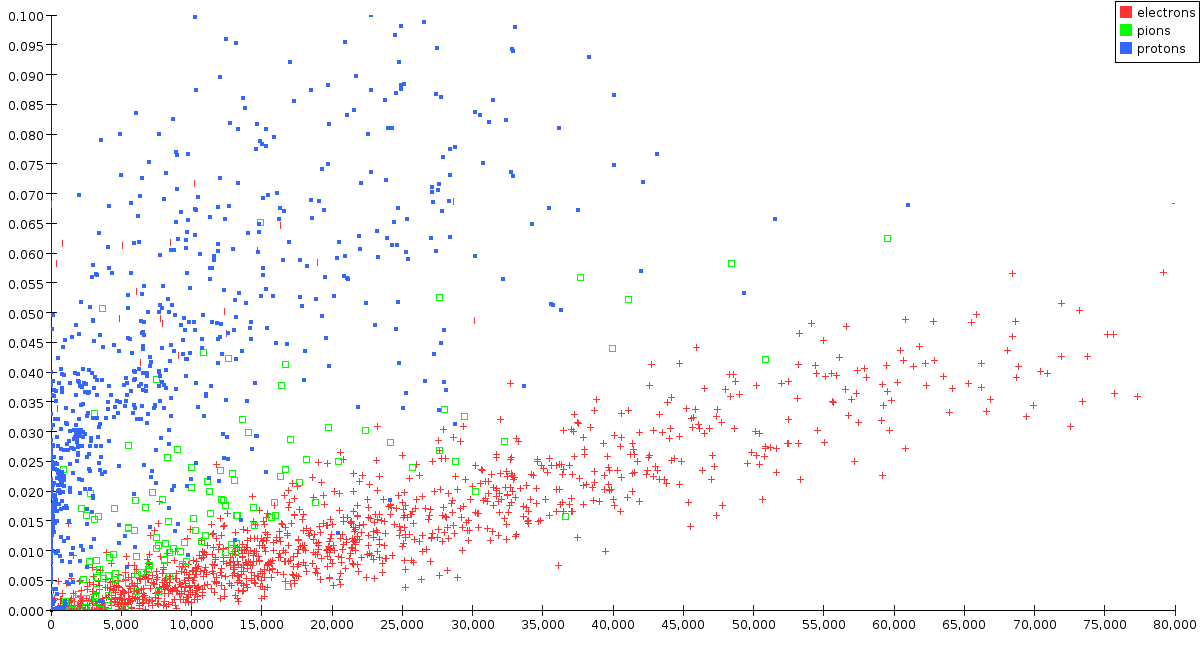}
\caption{\label{fig:Plot-of-S} Correlation between scintillator (S) vs. Čerenkov (C) signals (arbitrary units) in ADRIANO2 dual-readout calorimeter.
{(top)} Neutral particles where photons are marked in red, and neutrons in black).
{(bottom)} Charged particles with electrons in red, pions in green, protons in blue.
Electromagnetic particles are from simulated process $\eta\rightarrow\gamma A'_{17}\rightarrow\gamma e^{+}e^{-}$, while hadrons are from QCD background simulated with GenieHad.}
\end{figure}

One major advantage of dual- and triple-readout techniques is their ability to perform event-by-event particle identification by comparing the different signal components. Figure~\ref{fig:Plot-of-S} shows the separation between particle species for a typical BSM process (e.g., $\eta\rightarrow\gamma A'_{17}\rightarrow\gamma e^{+}e^{-}$) versus QCD background. The top plot shows $>4\sigma$ separation between photons and neutrons, well exceeding the 99\% PID efficiency requirement. For comparison, a conventional calorimeter with single readout measuring $dE/dx$ would only achieve the resolution shown by the vertical projection, resulting in significant degradation of PID performance, especially  for photons and neutrons.

For hadronic showers, the multiple-readout system also enables event-by-event compensation, reducing fluctuations in hadronic response and improving energy resolution.
In the ADRIANO3 region, the third readout is implemented using thin RPCs with $^{10}$Gd-doped glass. These detectors are highly efficient at detecting gamma quanta produced via neutron capture reactions $(n,\gamma)$ processes, followed by $\gamma$ emission and internal conversion: 
\[^{155}\text{Gd} + ^{1}_{0}n \to\ ^{156}\text{Gd}^{*} \to ^{156}\text{Gd} + \gamma,\]
\[^{157}\text{Gd} + ^{1}_{0}n \to\ ^{158}\text{Gd}^{*} \to ^{158}\text{Gd} + \gamma.\]
The resulting $\gamma$ particles decay into $e^{+}e^{-}$ pairs, which subsequently ionize the RPC gas, providing a neutron-sensitive signal largely independent of electromagnetic or charged hadronic backgrounds. This RPC layer is the third component of the triple-readout calorimeter. Due to Gd's very large neutron capture cross section between $\sim 6\times 10^{4}$ barn and $\sim 2.5\times 10^{5}$ barn~\cite{Brown:2018jhj,NSE05-64,Plompen:2020due}, depending on the isotope, the probability of capturing a thermal neutron in a 5~mm glass doped with 5\% Gd is essentially 100\%.
Even at lower Gd concentrations (e.g., 1--2\%), the capture probability would still be very high (above 90\%). Higher-energy neutrons can be detected in the plastic scintillating tiles of ADRIANO3 using standard pulse shape discrimination techniques and the timing spectrum of the shower. Ongoing studies are optimizing the ADRIANO3 layout and evaluating its sensitivity to neutrons.

The high granularity of the calorimeter presents significant challenges for the front-end and readout electronics. The system must meet stringent requirements in terms of high dynamic range while maintaining low noise performance to ensure accurate signal reconstruction over the full energy spectrum. In addition, high-precision timing capabilities are essential to cope with the high interaction rate and support trigger decisions. At the same time, the electronics must operate with low power consumption, targeting approximately $\sim$25 mW/channel, in order to limit the overall thermal load and ensure stable operation. The readout electronics will be embedded within the detector structure, which is composed largely of lead-glass. Therefore, thermal management and power efficiency are critical. Additionally, both scintillation and Čerenkov signals must be available to the Level-0 trigger system, requiring a low-latency architecture.

High-performance ASICs developed for the HL-LHC upgrades~\cite{CMS:2017lum,ATLAS:2017svb}, such as HGCROC3~\cite{Bouyjou:2022hfa} and TOFHIR2~\cite{Albuquerque:2020uyy}, are under active evaluation for REDTOP. These chips offer excellent time resolution (25--40 ps) and power consumption within the target range ($\sim$20 mW/channel). Although originally designed for higher light yields than those expected from ADRIANO, they can be adapted through careful tuning of the internal preamplifier gain stages.

A key challenge with adopting HGCROC3 for ADRIANO2 and ADRIANO3 is its design specification, such as the continuous readout at the LHC beam crossing frequency (40 MHz). This is approximately one order of magnitude slower than the inelastic event rate at REDTOP. However, the event topology at an $\eta/\eta^{\prime}$ factory like REDTOP is fundamentally different from that of the LHC. While the HL-LHC detectors experience an average event multiplicity of $\sim10^4$ particles per bunch crossing, REDTOP events involve only $\sim$6-7 final-state particles on average, resulting in a low detector occupancy of approximately $3.5 \times 10^{-4}$.

To take advantage of this, REDTOP will implement a Region of Interest (ROI)-based readout scheme, which means that only those ROIs that contain activity will be read out. Thanks to the low multiplicity, this strategy dramatically reduces the effective readout rate per channel, allowing it to match the 40 MHz limit of chips like HGCROC3, despite the higher global event rate. Studies are currently underway to finalize the readout architecture and optimize data throughput, latency, and power usage while ensuring compatibility with triggering and physics performance requirements.
\end{sloppypar}

\subsection{Superconducting Solenoid}
\begin{sloppypar}
A uniform magnetic field is an important element of the REDTOP detector, enabling precise momentum reconstruction of charged particles emerging from the target region and traversing the tracking system. The solenoidal configuration has been selected to provide a homogeneous field over the tracking volume while minimizing material and geometrical constraints on the surrounding subsystems. A solenoidal magnetic field of 0.6~T is required to measure the transverse momentum ($P_{t}$) of charged particles. In addition to providing momentum measurement, the magnetic field serves to magnetically contain low-momentum particles, preventing them from reaching the calorimeter and reducing background. The solenoid previously used in the now-dismantled FINUDA experiment~\citep{Bertani:1999ve} meets all operational and dimensional specifications required by REDTOP. The solenoid will be operated at approximately half its maximum field strength, significantly reducing the amount of iron required for the return yoke. 
\end{sloppypar}

\subsection{The Event Trigger System}\label{sec:trigger_system}
\begin{sloppypar}
To meet the goal of producing $5 \times10^{13}$ $\eta$ mesons per year (or $5\times10^{6}$ $\eta$/s, assuming $10^{7}$ seconds of useful beam time), the experiment requires an inelastic $p$-target interaction rate of approximately $7\times10^{8}$. 
This rate can be achieved with a proton beam intensity of $10^{11}$ $p$/s  and a Li or Be target with an effective thickness of about $2\times10^{-2}$ nuclear interaction lengths (7.7 mm for Li, 2.3 mm for Be), eventually divided into thinner layers. The resulting inelastic collision rate will produce a large flux of particles in the detector, predominantly hadrons generated by proton--nucleus interactions. The average particle multiplicity per collision is approximately 3.5 (charged) and 1.8 (neutral) for Li targets (3.8 and 1.9 for Be). This event rate exceeds that observed in all previous fixed-target experiments by more than an order of magnitude~\cite{Belyaev:2021cyr}, requiring the use of ultra-fast detectors and a multi-level trigger system, including topological filtering strategies, to reduce data volume while maintaining sensitivity to rare physics signatures.

REDTOP will implement a three-level trigger system designed to reduce the raw input rate by a factor of over 2,300, sufficient to limit data storage to a few petabytes per year. The trigger system leverages technologies developed for other high-rate experiments, reducing both costs and technical risks. The Level-0 (L0) trigger selects events based on fast, global features using low-latency data from the ADRIANO2 and ADRIANO3 calorimeters and the Čerenkov-TOF system. 
It aims to reject uninteresting events within a few tens of nanoseconds. Events of interest, primarily $\eta/\eta^{\prime}$ decays, typically include at least two leptons or pions and, in most cases, one or more energetic photons. In contrast, background hadrons from beam-target interactions are generally slower and less energetic, in most cases falling below the Čerenkov threshold of the Čerenkov-TOF.
The L0 trigger logic includes:
\begin{itemize}
    \item a minimum integrated Čerenkov signal from ADRIANO2,
    \item at least two Čerenkov-TOF clusters with energy above the threshold and TOF below a defined value,
    \item a scintillation to Čerenkov signal ratio in ADRIANO2 consistent with electromagnetic showers.
\end{itemize}
The above selection criteria constitute the baseline L0 trigger and have been validated through realistic simulations.  
Simulations indicate that this selection strategy will result in an event rate reduction factor of $\sim$4.6 at the L0 stage. 
After digitization and data compression, the average event size is estimated to be approximately 1.5~kB. 
Under the assumption of 12-bit charge, time digitization, and 18-bit channel addressing at a sampling rate of 700~MHz, the resulting data rate entering the Level-1 (L1) stage is estimated to be about 230~Gbps. This bandwidth can be handled using a few hundred optical links. The L0 decision logic  is fully pipelined and tolerant of latency. Although events arrive, on average, every $\sim$1.4 ns, the allowed processing time per event can be extended to several hundred nanoseconds.

The Level-1 (L1) trigger performs refined event selection based on pattern recognition and localized detector information, with the primary objective of suppressing background from slow baryons and photon conversions. Given the relatively low occupancy in the tracking detectors, the implementation of L1 logic using Vertically Integrated Pattern Recognition Associative Memory (VIPRAM) \cite{Deputch:2011zz}, represents a promising hardware solution.
The L1 selection strategy combines tracking, timing, and calorimetric observables. The criteria applied to suppress baryon induced background are:  
\begin{itemize}
    \item at least three clusters in the central tracker, each separated by at least 4.5 mm,
    \item at least two proto-tracks with three LGAD clusters and one Čerenkov-TOF cluster, consistent with opposite charge and fast TOF,
    \item Čerenkov-TOF energy for both tracks is above threshold,
    \item at least three ADRIANO2 clusters with EM-like signal ratios.
\end{itemize}
Complementary topological criteria are applied to suppress photon conversion background:
\begin{itemize}
    \item no vertex detector clusters are associated with close (<3 mm) oppositely charged tracks in  $r-\phi$ and z.
    \item no central tracker clusters are present with oppositely charged tracks separated by less than 4.5 mm.
\end{itemize}

The effectiveness of this selection strategy has been evaluated using simulations, which indicate an event rate  reduction factor of $\sim$60 at L1, lowering the data rate to $\sim$3.8 Gbps. A key technical challenge is the development of dedicated L1 processors capable of reconstructing proto-tracks and applying selection logic at rates of up to 100 kHz. This requirement can be met using a massively parallel FPGA based architecture combined with time-multiplexing. In a round-robin distribution scheme with a multiplexing factor of 10, each processor would have $\sim 5 \mu s$  per event, which is sufficient for L1 logic. Furthermore, the possible implementation of AI/ML techniques in the L1 trigger is currently under consideration as a potential enhancement of pattern recognition performance.
\begin{table*}[t]
\centering
\renewcommand{\arraystretch}{1.2}
\begin{tabular}{ccccc}
\toprule 
{Trigger stage} & 
{Input event rate (Hz)} & 
{Event size (bytes)} & 
{Input data rate (bytes/s)} & 
{Event (rejection)}\\
\midrule 
{Level 0} & {$7.\times10^{8}$} & {$1.4\times10^{3}$} & {$9.8\times10^{11}$} & {$\sim4.6$}\\
{Level 1} & {$1.5\times10^{8}$} & {$1.5\times10^{3}$} & {$2.3\times10^{11}$} & {$\sim60$}\\
{Level 2} & {$2.5\times10^{6}$} & {$1.5\times10^{3}$} & {$3.8\times10^{9}$} & {$\sim4.5$}\\
{Storage} & {$0.56\times10^{6}$} & {$1.6\times10^{3}$} & {$0.9\times10^{9}$} & \\ \bottomrule
\end{tabular}
\caption{Summary of event and data rates at the successive stage of the REDTOP trigger and data acquisition chain. The chain reduces the initial $\sim 7 \times 10^{8}$~Hz interaction rate to a sustained storage rate of $\sim 0.56 \times 10^{6}$~Hz, corresponding to a total event rejection of $\sim 1250$ and a final data flow of $\sim 0.9$~GB/s.}
\label{table:triggerLevels}
\end{table*}

The Level-2 (L2) trigger performs event-level reconstruction and the final physics-based selection using the full detector readout. 
At this stage, the processing includes the identification of the target foil in which the primary interaction occurred, approximate track and calorimeter shower reconstruction, and the identification of possible detached secondary vertices.
In the L2 trigger, events are accepted based on signal topology characteristics consistent with the $\eta$ and $\eta^{\prime}$ decays. 
These include final states with two fully identified leptons, two oppositely charged pions accompanied by two calorimetric showers, and four-pion final states.
Furthermore, events containing any two oppositely charged tracks that form a secondary vertex located more than 5 cm from the interaction point are selected as triggering signatures of long-lived particle decays. Trigger L2 is implemented entirely in software and executed on a CPU farm. The expected input event rate at this stage is approximately 2.5~MHz, corresponding to data of about 3.8 GB/s. Assuming an average processing time of less than 100~ms/event, a computing farm of 2,000 CPUs would be sufficient to perform the above tasks.

The REDTOP trigger system is designed to reduce the raw inelastic interaction rate of $\sim7\times10^{8}$ down to a manageable rate of a few MHz of events for permanent storage. Assuming an average final event size of $\sim$1.6 kB, this corresponds to an output data rate of about 0.9 Gbps, or roughly 9 PB/year, which is well within the capabilities of modern data handling infrastructure. This overall reduction factor of approximately 2,300 is achieved through a combination of digitization and data compression at the detector front-end, together with the three trigger levels described above, which perform progressive event reconstruction and filtering (see Table~\ref{table:triggerLevels}). Further developments may include topological triggers at the L0 stage, replacing or complementing the global trigger used in the present study, which could further improve rejection rates.

Together, these subsystems form an integrated detector with near-$4\pi$ coverage and excellent sensitivity to low-energy leptons and photons, features that are essential for reconstructing the full kinematics of $\eta$ and $\eta^{\prime}$ decays. 
The resulting geometry provides uniform acceptance across all channels of interest, enabling precise measurements of symmetry violating observables, rare decay topologies, and potential signals of hidden-sector particles. 
The detector concept shown in Fig.~\ref{fig:reddetector} represents the baseline configuration used throughout the simulation studies presented in the following sections. All physics sensitivity estimates, including those for $CP$-violating observables, rare-portal-induced decays, and nonperturbative QCD processes, are derived under the realistic assumptions implemented in this design. 
As such, the conceptual design described above serves as the reference configuration for evaluating the expected performance of REDTOP and for guiding the ongoing optimization of its subsystems.
\end{sloppypar}

\section{Physics Sensitivity Studies}\label{PhysicsSensitivity}
In this section, we present detailed sensitivity studies for a representative set of $\eta$ and $\eta^{\prime}$ decay channels that drive the physics case for REDTOP.  
All results are based on full detector simulations incorporating realistic assumptions on geometry, material budget, beam conditions, and background composition.  
The selected channels span the main categories discussed in the Introduction (Sec.~\ref{sec:Intro}) and include probes of discrete symmetries, searches for light hidden-sector states, and tests of lepton-flavor universality. These studies quantify the achievable precision for each observable and define the detector and beam requirements needed to fully exploit the experiment’s potential.

\subsection{Searches for New Particles and Fields }\label{SectionFramework}
Theoretical frameworks addressing light and weakly coupled new particles often introduce a \textit{hidden sector} that interacts with the Standard Model (SM) through a limited set of renormalizable operators known as \textit{portals}~\cite{Batell:2009di}. Portals provide the simplest gauge-invariant mechanisms for communication between visible and hidden degrees of freedom. They involve new fields such as vector bosons, scalars, pseudoscalars, or heavy neutral leptons that couple to the SM via gauge-singlet operators of mass dimension four or less~\cite{Patt:2006fw,Arkani-Hamed:2008hhe,Jaeckel:2010ni}. Such renormalizable operators are characterized by small, dimensionless couplings and may lead to observable effects for new states in the MeV-GeV mass range~\cite{Knapen:2017xzo}. In addition, higher-dimensional operators, as present, for example, in axion models, provide further opportunities for exploring phenomena beyond the SM at suppressed levels but potentially detectable at a super-$\eta/\eta^\prime$ factory.

\begin{sloppypar}
In this context, investigating processes involving particles that are neutral under all Standard Model charges is particularly relevant for Light Dark Matter searches. In line with the above considerations, the $\eta$ and $\eta^{\prime}$ mesons stand out as particularly relevant for such studies. Both are members of the ground-state pseudoscalar nonet and play an important role in understanding low-energy QCD~\cite{Gasser:1983yg,Gasser:1984gg,Feldmann:1999uf}. As already remarked in the Introduction (Sec.~\ref{sec:Intro}), both have isospin and angular momentum equal to zero, negative parity, and charge conjugation equal to $+1$ [$I^G(J^{PC}) = 0^+(0^{-+})$]~\cite{ParticleDataGroup:2024cfk}. 
From an experimental perspective, the absence of SM charges, and thus of charged currents, suppresses background processes, while electromagnetic and strong decays occur only at the $\mathcal{O}(10^{-6})$ level, relative to typical hadronic widths (e.g., $\Gamma_\eta = 1.3$~keV) further enhancing the visibility of rare processes and potential BSM signals.
\end{sloppypar}

A high-intensity $\eta/\eta^{\prime}$ factory offers an almost unique opportunity shared only with the SHIP experiment~\cite{SHiP:2021nfo} at CERN and HHAS~\cite{Chen:2024wad} at HIAF to probe all four theoretical portals that could mediate interactions between the dark sector and the Standard Model. To investigate these portals and their role in accessing hidden sectors, several $\eta$ and $\eta^{\prime}$ decay processes have been identified as particularly promising probes of these new interactions. The expected sensitivities to each portal, derived from detailed simulations of representative processes, are summarized in the following sections.

\subsubsection{The Vector Portal}
\begin{sloppypar}
The vector portal encompasses a wide class of theoretical frameworks in which a new vector boson mediates interactions between the Standard Model (SM) and hidden sectors. Representative examples include the minimal dark photon model~\cite{Holdom:1985ag}, the leptophobic $B$-boson model~\cite{Tulin:2014tya,Escribano:2018cwg,Escribano:2022njt,Escribano:2025yoq}, and the protophobic fifth-force scenario~\cite{Feng:2016jff,Feng:2016ysn}. Within this framework, the $\eta$ and $\eta^{\prime}$ mesons can probe vector mediated interactions through radiative decays of the type $\eta\!\to\!\gamma A^{\prime}$, where the new vector boson $A^{\prime}$ subsequently decays depending on its coupling structure: vectors with isovector couplings decay primarily to $\ell^+\ell^-$ or $\pi^+\pi^-$, while vectors with isoscalar couplings decay to $\pi^0 \gamma$ at low masses or to $\pi^+\pi^-\pi^0$ at higher masses~\cite{Tulin:2014tya}. This makes the channel $\eta \to \pi^0 \gamma \gamma$ (via $\eta \to \gamma A^{\prime}$, $A^{\prime} \to \pi^0 \gamma$) a particularly sensitive probe of isoscalar vector mediators.

In the minimal dark-photon model, the SM is extended by a single new state $A^{\prime}$ that couples to visible matter via a kinetic-mixing parameter $\varepsilon$~\cite{Holdom:1985ag}. This mixing induces small effective couplings of $A^{\prime}$ to charged particles, leading to potentially observable decays 
$\eta/\eta^{\prime}\!\rightarrow\!\gamma A^{\prime}\!\rightarrow\!\gamma e^{+}e^{-}$ and 
$\eta/\eta^{\prime}\!\rightarrow\!\gamma A^{\prime}\!\rightarrow\!\gamma\mu^{+}\mu^{-}$.  
REDTOP, with $\eta/\eta^{\prime}$ produced samples exceeding $10^{14}$ mesons, is expected to be sensitive to branching fractions well below $10^{-9}$ for these modes, once reconstruction efficiency is taken into account.

\begin{figure}[t]
\centering
\includegraphics[width=0.90\columnwidth]{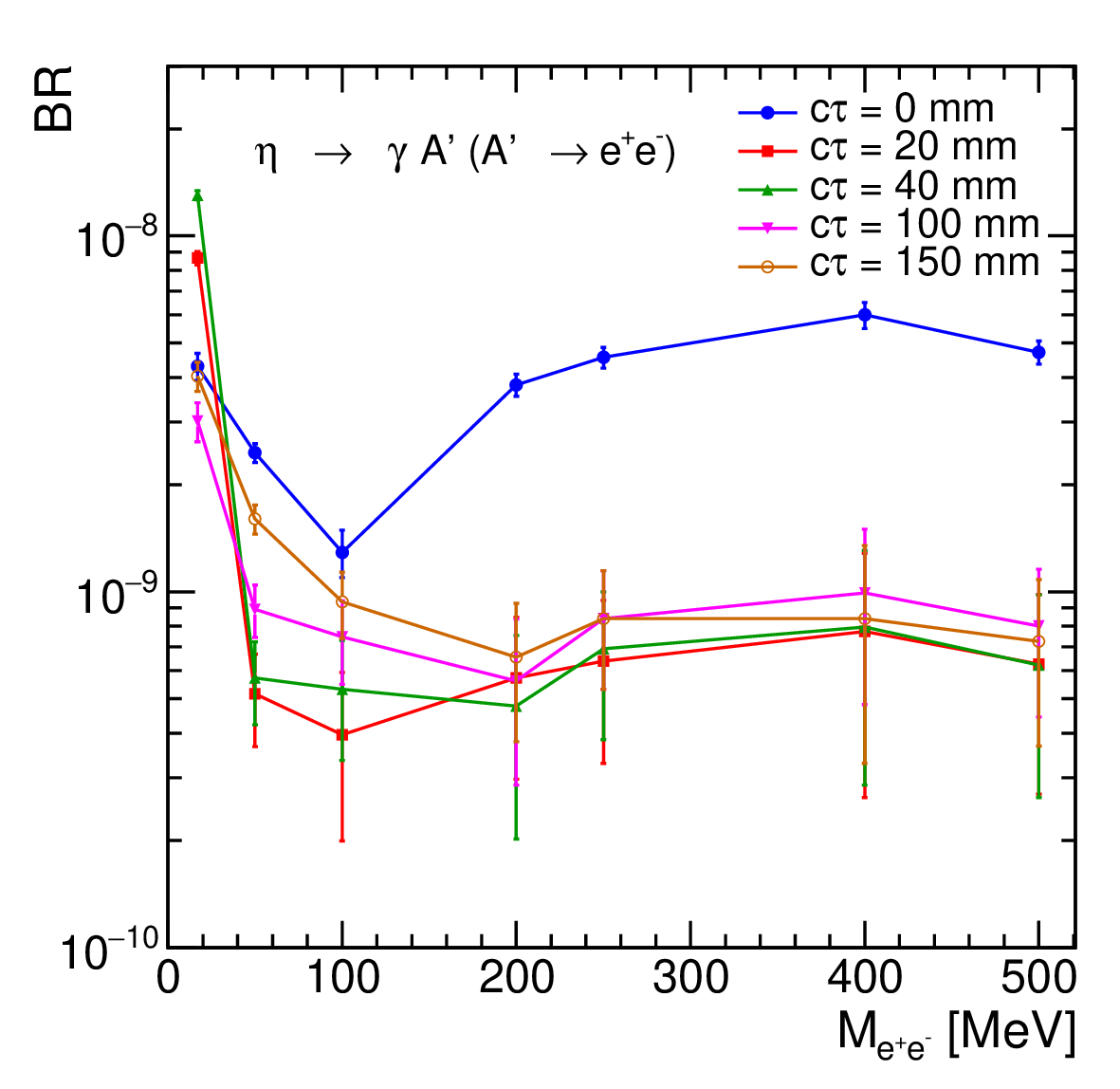} 
\includegraphics[width=0.90\columnwidth]{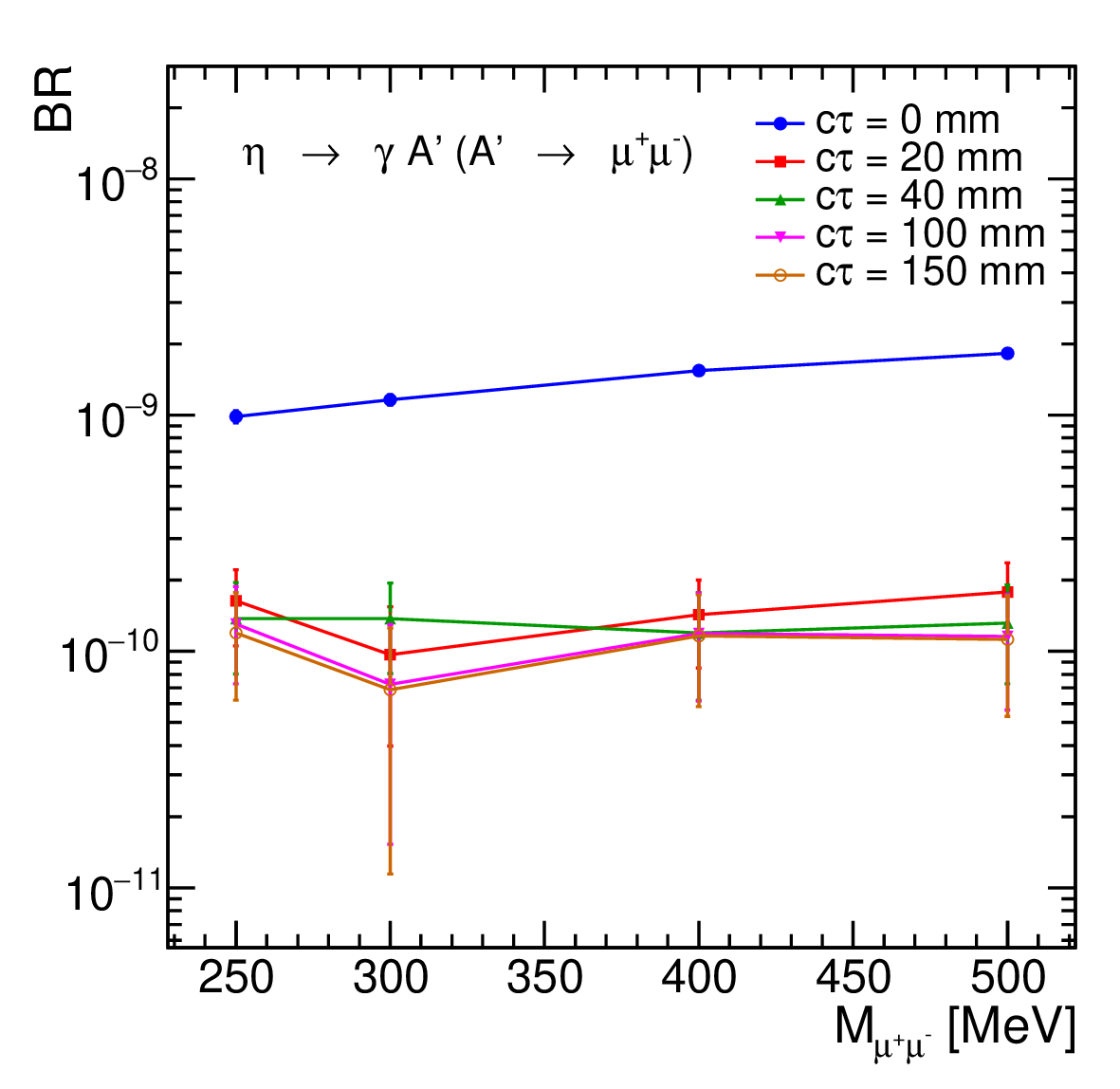}
\caption{Branching ratio sensitivity for the process {(upper)}
$\eta\rightarrow\gamma A'$, $A'\rightarrow e^{+}e^{-}$  and {(lower)} $\eta\rightarrow\gamma A'$, $A'\rightarrow\mu^{+}\mu^{-}$
as a function of the mass and $c\tau$ of a long-lived vector boson $A'$. 
}
\label{fig:eta2gammaAp_ee_br-vtx}
\end{figure}

To quantify the REDTOP reach in this model, two dedicated simulation studies were performed to evaluate the sensitivity to vector mediated decays and to assess the performance of key detector components.  
The first, a \emph{bump-hunt} analysis, assumes a promptly decaying $A^{\prime}$ with a decay length too short to be resolved by the tracking system.  
The second, a \emph{detached-vertex} analysis, targets long-lived mediators that decay at a measurable distance from the $\eta$ production point.  
The resulting branching ratio sensitivities are shown in Fig.~\ref{fig:eta2gammaAp_ee_br-vtx} as a function of the assumed $A^{\prime}$ mass.  
Results are presented for two final-state topologies: (upper panel) $A^{\prime}\!\to\! e^{+}e^{-}$ and (lower panel) $A^{\prime}\!\to\!\mu^{+}\mu^{-}$.  
The error bars represent the statistical uncertainties on the simulated data points.

\begin{figure}[t] 
\centering
\includegraphics[width=0.95\columnwidth]{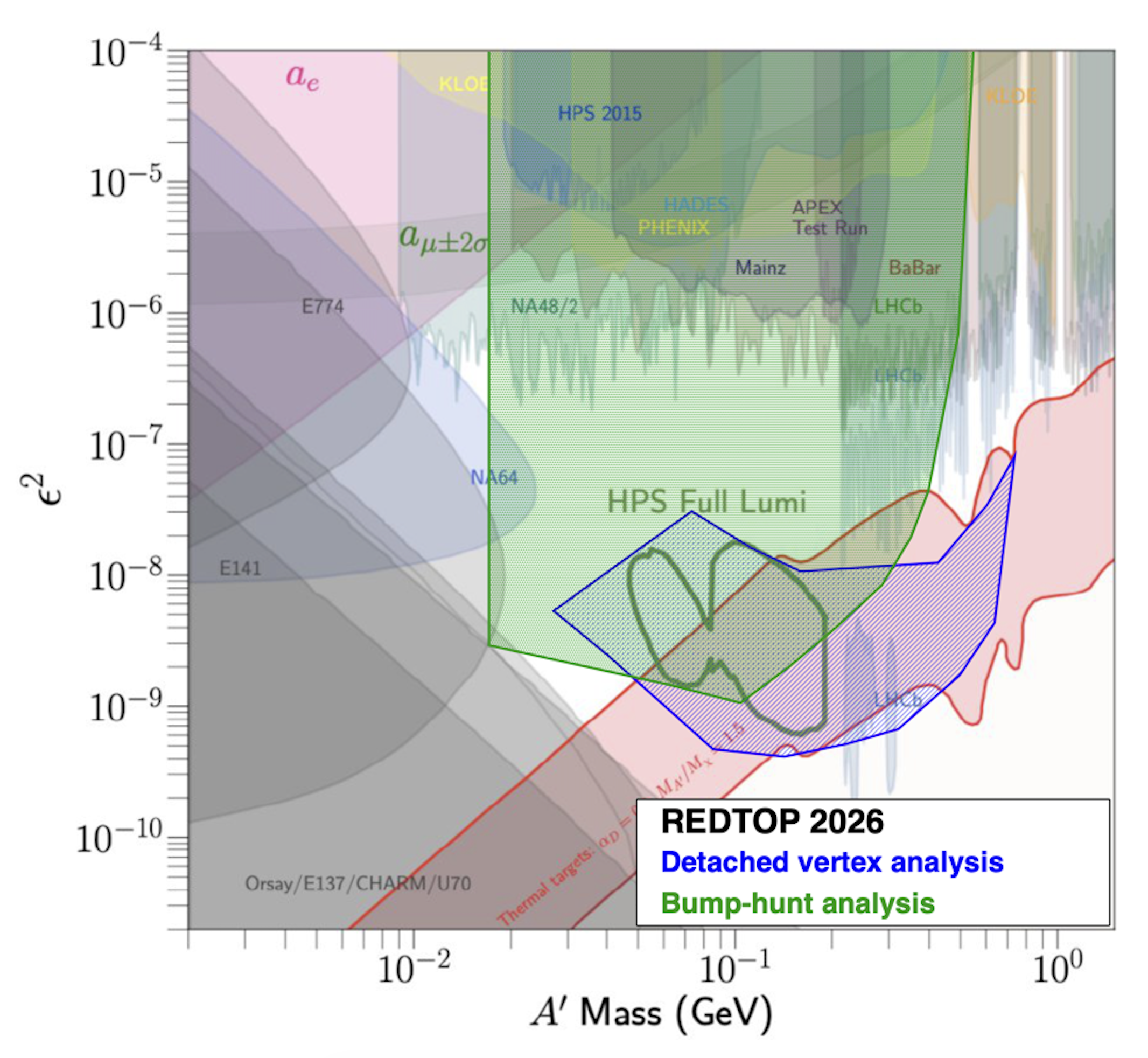}
\caption{Projected REDTOP sensitivity to the kinetic-mixing parameter $\varepsilon^{2}$ 
for the process $\eta\!\to\!\gamma A'$, assuming an integrated beam flux of $3.3\times10^{18}$~POT (Protons On Target). 
Results from the {(green area)}  \emph{bump-hunt} analysis, and {(blue area)} results from the \emph{detached-vertex} analysis.}
\label{fig:eta2gammaAp_eeepsilon}
\end{figure}

From these studies, the corresponding sensitivity to the kinetic-mixing parameter $\varepsilon^{2}$ was derived and is shown in Fig.~\ref{fig:eta2gammaAp_eeepsilon} for both the (green area) \emph{bump-hunt} and (blue area) \emph{detached-vertex} analyses, superimposed on the existing constraints from previous measurements.  
The projected REDTOP performance extends the experimental reach over most of the presently unexplored $\varepsilon^{2}$ parameter space, providing competitive or superior sensitivity across a broad mass range. 
The sensitivity curves correspond to an integrated yield of $10^{14}$ $\eta$ mesons, equivalent to a three year data taking period under the beam and detector conditions described in Sec.~\ref{theDetector}.
\end{sloppypar}

\subsubsection{The Scalar Portal}
\begin{sloppypar}
The scalar portal connects the Standard Model to a hidden sector through a light scalar field $H$ that potentially mixes with the SM Higgs boson. Such a mediator can be produced in $\eta$ decays involving a neutral pion, for example, $\eta\!\to\!\pi^{0}H$ followed by $H\!\to\! \ell^{+}\ell^{-}$ or $H\!\to\!\pi^{+}\pi^{-}$.  
Three representative scalar portal models have been considered for the studies presented here: the Minimal Dark-Scalar Model, the Spontaneous Flavor-Violation Model~\cite{Egana-Ugrinovic:2018znw}, and the Flavor-Specific Scalar Model, which yield similar experimental signatures and are closely related to the Two-Higgs-Doublet scenario~\cite{Batell:2021xsi,Abdallah:2020vgg}. As emphasized in Ref.~\cite{Batell:2018fqo}, REDTOP is particularly sensitive to scalars coupling dominantly to first-generation quarks.     

In the Spontaneous Flavor-Violation framework, the scalar $H$ couples primarily to up quarks, the so-called \emph{hadrophilic scalar mediator}.  
Its production proceeds via $\eta\rightarrow\pi^0 H$, with a branching ratio given by~\cite{Batell:2018fqo} 
\begin{eqnarray}
{\mathrm{Br}}(\eta \rightarrow \pi^0 H) = \frac{c_{H\pi^0\eta}^2\, g_u^2\, B^2} {16 \pi m_\eta \Gamma_\eta}\,
\lambda^{1/2}\!\left(1, \frac{m_H^2}{m_\eta^2},\frac{m_{\pi^0}^2}{m_\eta^2}\right),
\label{eq:Breta-Pi-S} 
\end{eqnarray}
where $\lambda(a,b,c)=a^2+b^2+c^2-2ab-2ac-2bc$ is the triangle function~\cite{Byckling:1971vca}, 
$B\simeq m_\pi^2/(m_u+m_d)\approx2.6$~GeV, 
and $c_{S\pi^0\eta}=\tfrac{1}{\sqrt{3}}\cos\theta-\sqrt{\tfrac{2}{3}}\sin\theta$ 
accounts for $\eta$--$\eta^{\prime}$ mixing with $\theta\approx-20^\circ$~\cite{Feldmann:1999uf}.

\begin{figure}[t]
\centering
\includegraphics[width=0.9\columnwidth]{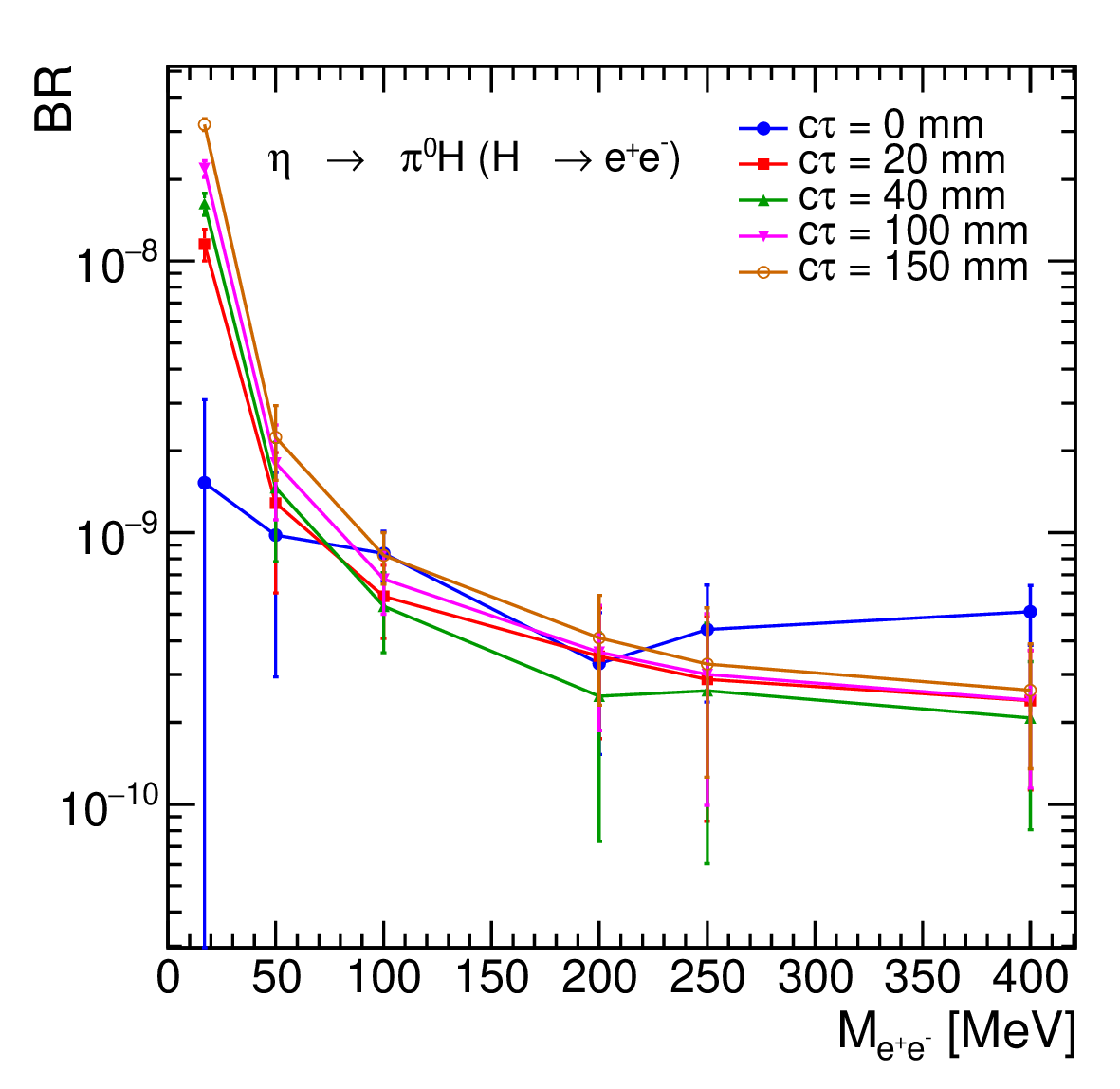} 
\includegraphics[width=0.9\columnwidth]{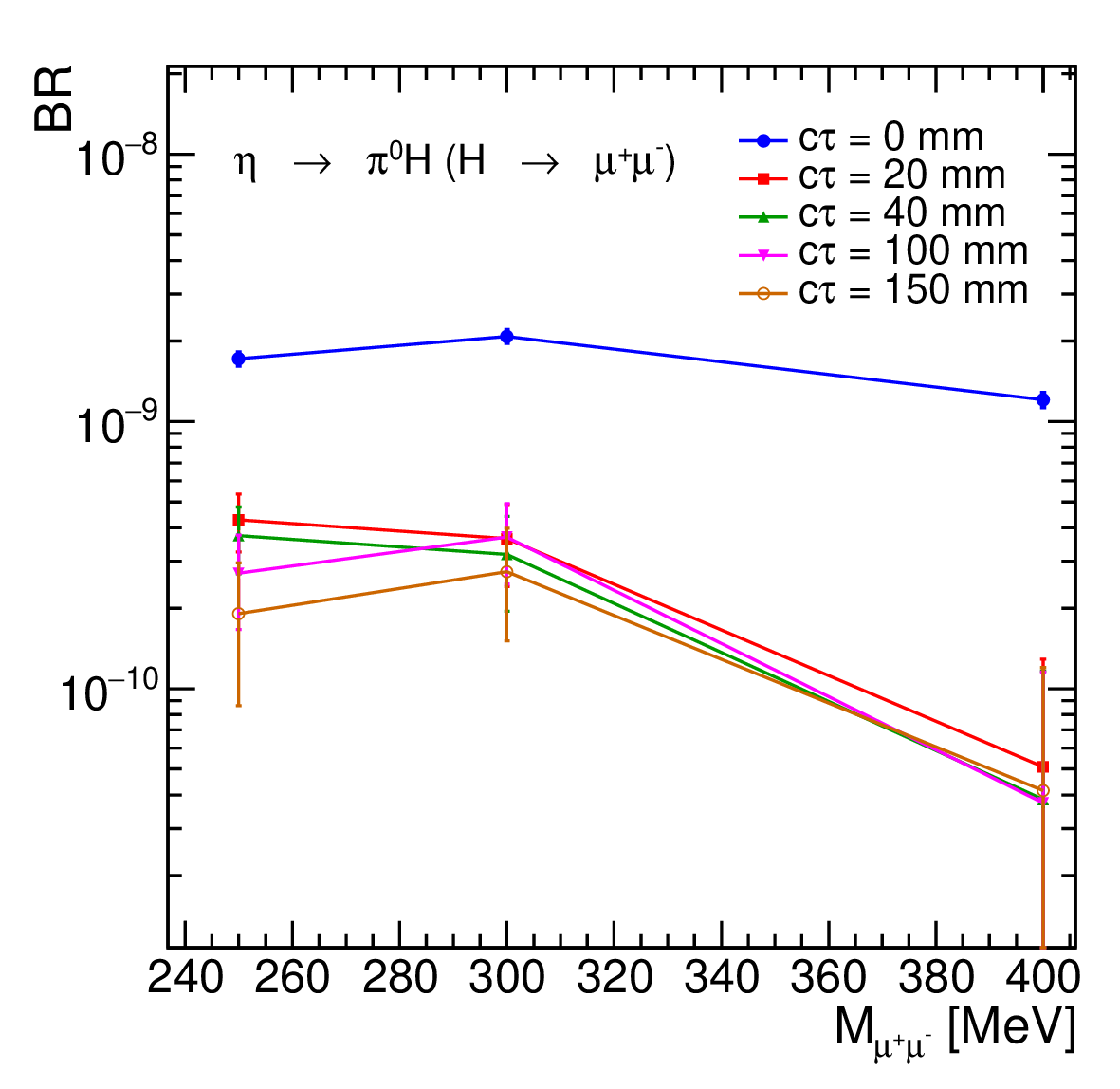}
\caption{Branching ratio sensitivity for the process {(upper)}  $\eta\rightarrow\pi^{0}H$, $H\rightarrow e^{+}e^{-}$
and {(lower)}  $\eta\rightarrow\pi^{0}H$, $H\rightarrow\mu^{+}\mu^{-}$ as a function of the mass and $c\tau$ of a long-lived scalar boson $H$.}
\label{fig:eta2pi0H_ee_br-vtx}
\end{figure}

To evaluate the experimental reach, two complementary analyses were performed, analogous to those used for the vector portal. The \emph{bump-hunt} analysis targets promptly decaying scalars, while the \emph{detached-vertex} analysis probes scenarios with measurable decay lengths. The resulting branching ratio sensitivities are shown in Fig.~\ref{fig:eta2pi0H_ee_br-vtx} as functions of the scalar boson mass and lifetime $c\tau$. The upper panel corresponds to $H\!\to\! e^{+}e^{-}$ decays, while the lower panel shows the results for $H\!\to\! \mu^{+}\mu^{-}$.
The error bars represent the statistical uncertainties on the simulated data points.

\begin{figure}[t]
\centering
\includegraphics[scale=0.4]{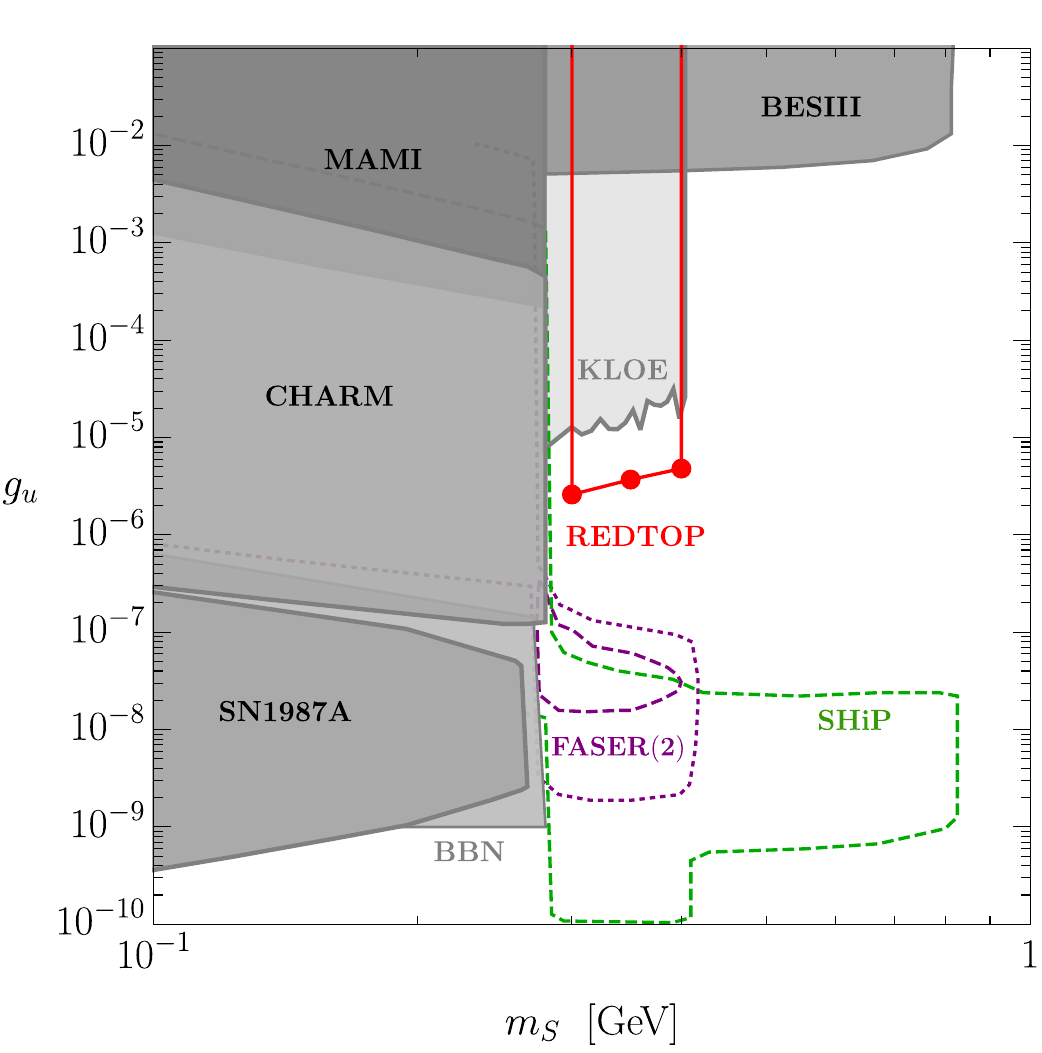}
\caption{REDTOP sensitivity to the the \emph{hadrophilic scalar mediator}, in the $\eta \rightarrow \pi^0 H$, $H\rightarrow \pi^+ \pi^-$ mode. 
We display the projected sensitivity of REDTOP (red line) in the $m_H-g_u$ plane a the bump hunt analysis based on $3.3\times{10}^{18}$ POT for the three mass points $m_H = {300, 350, 400}$ MeV. Also shown are various existing constraints and projections from other planned or proposed experiments described in Ref.~\cite{Batell:2018fqo,Kling:2021fwx}.}
\label{fig:dark_scalar}
\end{figure}

For the parameter region relevant to REDTOP, the scalar is expected to decay promptly. The projected sensitivity to the coupling $g_u$ is shown in Fig.~\ref{fig:dark_scalar}, where the REDTOP reach is compared with existing constraints.  
The results for the subsequent decay channel $\eta\rightarrow \pi^0 H$ and $H\rightarrow \pi^+ \pi^-$ indicate sensitivity to couplings of order $g_u \sim \mathrm{few}\times10^{-6}$, extending the reach by a factor of several beyond recast bounds from KLOE~\cite{KLOE-2:2016zfv,Batell:2018fqo}.  
REDTOP also complements the projected coverage of long-lived particle searches at FASER, FASER2, and SHIP~\cite{Batell:2018fqo,Kling:2021fwx}, 
which are sensitive to longer lifetimes and smaller couplings. 
\end{sloppypar}

\subsubsection{The Pseudoscalar Portal}
\begin{sloppypar}
The pseudoscalar portal offers a particularly rich framework for exploring physics Beyond the Standard Model (BSM), with several recent theoretical models predicting the existence of new light pseudoscalar states. Among the most compelling ideas is the long-sought but still unconfirmed Peccei--Quinn mechanism, which introduces the so-called \textit{QCD axion}~\cite{Peccei:1977hh,Peccei:1977ur} originally proposed to solve the strong $CP$ problem in Quantum Chromodynamics (QCD)~\cite{Kim:2008hd}. Furthermore, Axion-Like Particles (ALPs)~\cite{DiLuzio:2020wdo} represent another important class of models related to the Pseudoscalar Portal. In addition to its theoretical elegance, ALPs~\cite{Bauer:2021wjo} have gained renewed interest in recent years due to their potential role in explaining anomalies observed in various experiments~\cite{Krasznahorkay:2015iga}. This has further motivated dedicated searches in rare $\eta$ and $\eta^{\prime}$ decays, where pseudoscalar signatures could manifest.

Of particular interest are axio-hadronic decays~\cite{Alves:2017avw} resulting in two charged or neutral pions. In these scenarios, GeV-scale dynamics coupling to the first generation of Standard Model fermions can produce a short lived QCD axion or ALP that decays predominantly into $e^+e^-$ pairs. Processes such as $\eta^{(\prime)} \to \pi\pi a$ therefore represent key experimental channels for probing pseudoscalar states that couple to quarks and gluons~\cite{Alves:2020xhf,Alves:2024dpa}. Such couplings would modify the topological vacuum structure of QCD~\cite{Callan:1976je} and contribute to the strong $CP$ phase, $\theta_{{QCD}}$~\cite{Callan:1977gz}. While an ALP might shift $\theta_{QCD}$ without resolving the $CP$ problem, potentially requiring further fine tuning, the discovery of a pseudoscalar that dynamically drives 
$\theta_{QCD} \to 0$ would provide compelling evidence for the existence of the QCD axion itself.

In this context, the processes foreseen for studies by REDTOP are:
\begin{itemize}
\item $p+\text{Li}\rightarrow\eta+X$ with $\eta\rightarrow\pi^{+}\pi^{-}a$ and \ $a\rightarrow e^{+}e^{-}$
\item $p+\text{Li}\rightarrow\eta+X$ with $\eta\rightarrow\pi^{0}\pi^{0}a$ and $a\rightarrow e^{+}e^{-}$
\item $p+\text{Li}\rightarrow\eta+X$ with $\eta\rightarrow\pi^{+}\pi^{-}a$ and \ $a\rightarrow\gamma\gamma$
\end{itemize}
As in previous cases, two distinct analyses were conducted to evaluate the detector performance: a \emph{bump-hunt} analysis to search for resonant mass peaks and a \emph{detached-vertex} analysis to probe long-lived pseudoscalar decays.
\begin{figure}[!t]
\centering 
\includegraphics*[width=0.87\columnwidth]{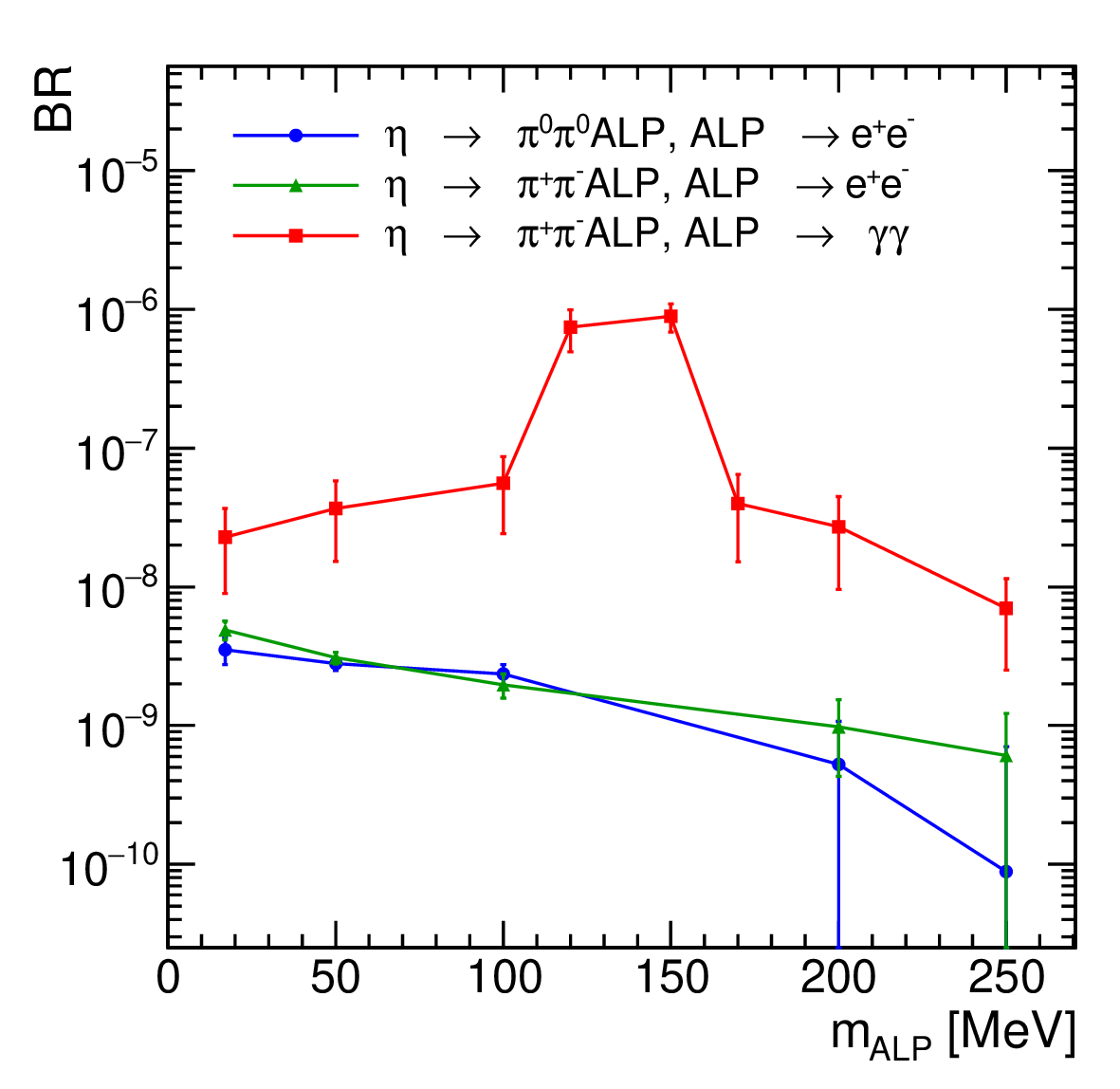} \\[-1.5mm]
\caption{Branching ratio sensitivity for the process: {(blue line)} $\eta\rightarrow\pi^{0}\pi^{0}a$ with $a\rightarrow e^{+}e^{-}$, {(green line)} $\eta\rightarrow\pi^{+}\pi^{-}a$ with $a\rightarrow e^{+}e^{-}$, and {(red line)} $\eta\rightarrow\pi^{+}\pi^{-}a$ with $a\rightarrow\gamma\gamma$ as a function of the pseudoscalar boson $a$ invariant mass.}
\label{fig:eta2pippimalp_ee_br}
\end{figure}
The resulting sensitivities for all considered processes are shown in Fig.~\ref{fig:eta2pippimalp_ee_br} as a function of the pseudoscalar boson \emph{a} mass.  
As expected, the process has a lower sensitivity for values of $M_{a}$ below 50~MeV due to the large contribution from photon conversions ($\gamma \to e^{+}e^{-}$) in the detector material. For higher mass values, the experiment can probe branching ratios down to $\mathcal{O}(10^{-9})$ at 90\% CL.

Two theoretical models associated with the pseudoscalar portal have been considered in the context of REDTOP studies: the \emph{piophobic axion model}, and axion-like particles (ALPs) characterized by either \emph{quark-dominated} or \emph{gluon-dominated} couplings. The piophobic axion model is particularly compelling, as it extends the viable mass range of the QCD axion into a region accessible to REDTOP sensitivity, offering a rare opportunity to experimentally probe this elusive solution to the strong $CP$ problem.
REDTOP has evaluated its sensitivity to this model by fitting the axion momentum spectrum using the theoretical matrix element presented in~\cite{Alves:2020xhf}. 
The $\chi^{2}$ probabilities from the fits are summarized in Table~\ref{table:qcdaxionfit} for three representative sets of model parameters. 
\begin{table}[t]
\centering
\renewcommand{\arraystretch}{1.2}
\begin{tabular}{cccc}
\toprule 
{POT} & $\eta\rightarrow\pi^{+}\pi^{-}a\;(a\rightarrow e^{+}e^{-})$ & BR  & $\chi^{2}$/ndof \\
 & Benchmark set & stat. error & \\
\midrule 
1.1$\times10{}^{13}$ & B1 & 1.5\% & 3.2\\ 
1.3$\times10{}^{13}$ & B2 & 1.4\% & 2.9\\ 
5.6$\times10{}^{12}$ & B3 & 2.1\% & 3.5\\
\bottomrule 
\end{tabular}\caption{Goodness of fit of the P$_{axion}$ distribution using the matrix element from Ref.~\cite{Alves:2020xhf} }
\label{table:qcdaxionfit}
\end{table}
A statistical error of less than  $<2\%$ across these fits demonstrates REDTOP excellent sensitivity to the \emph{piophobic axion model}.

\begin{figure}[t]
\centering
\includegraphics[width=0.90\columnwidth]{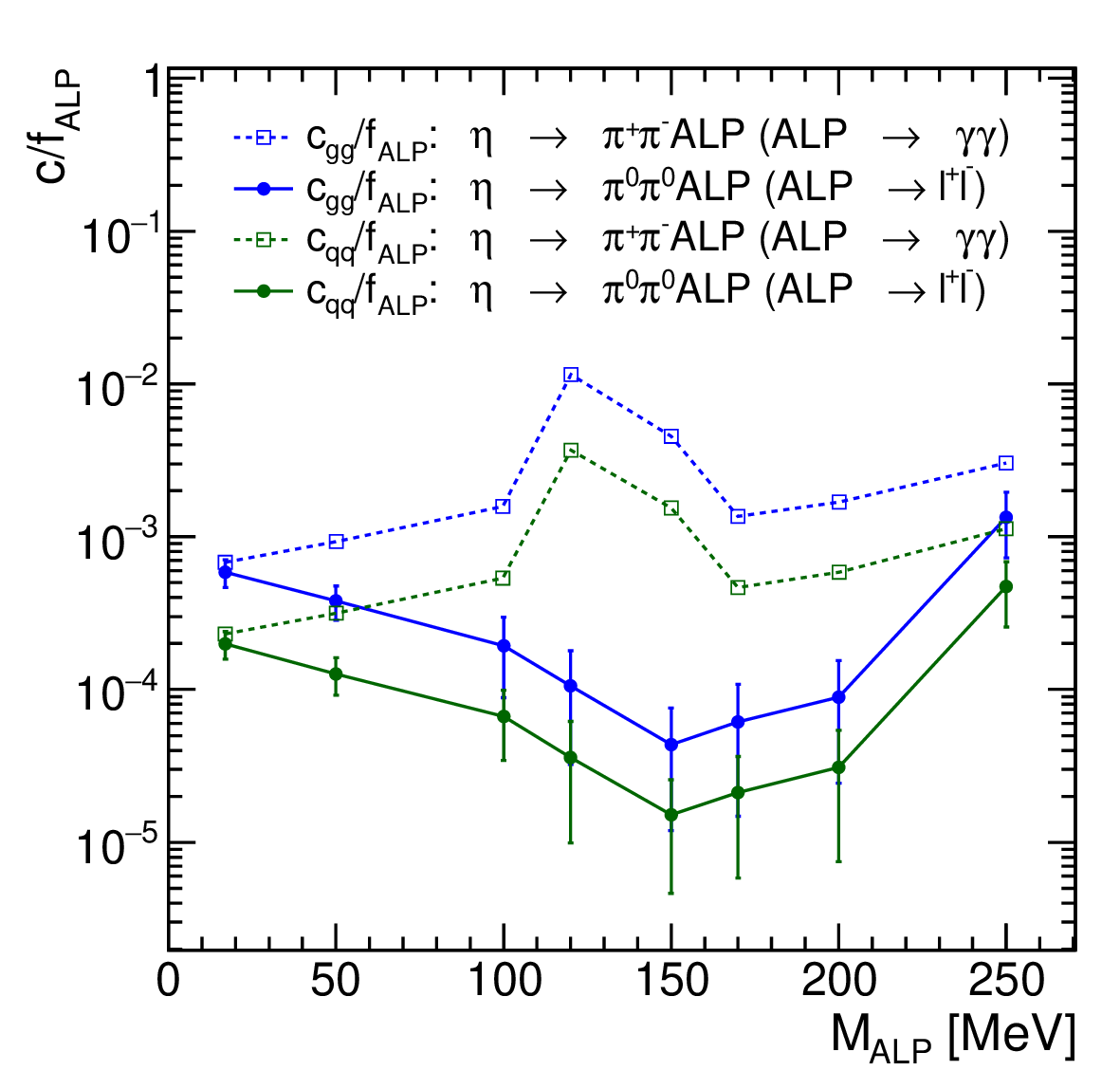}
\caption{Sensitivity to $c_{gg}/f_{a}$ (blue points) and $c_{qq}/f_{a}$ (green points) for the processes $\eta\rightarrow\pi^{+}\pi^{-}a$ and $\eta\rightarrow\pi^{0}\pi^{0}a$
as a function of the mass of a the ALP \emph{a}. The square symbols refer to the decay $a\rightarrow e^{+}e^{-}$ while the solid cirles are for the case: $a\rightarrow \gamma\gamma$. See text for details of the analysis.}
\label{fig:eta2pipiALP_cgg_cqq}
\end{figure}

Axion-Like Particles share the same types of interactions as the QCD axion, but differ in that they receive additional Peccei--Quinn symmetry breaking contributions to their masses. Consequently, ALP models exhibit a broader parameter space, as the ALP mass and decay constant are treated as independent parameters, unlike in the QCD axion scenario.
The key parameters in this formalism include the coupling constants $c_{gg}$, $c_{qq}$, and the ALP decay constant $f_{a}$, defined in~\cite{Alves:2020xhf}.
For the determination of REDTOP sensitivity to $c_{gg}$ and $c_{qq}$, we made the conservative assumption that \emph{a} only decays into Standard Model particles. In that case, the values of the coupling constants are also related to the width of \emph{a} and, consequently, to the capability of the detector to reconstruct detached vertices. Using the branching ratio values shown in  Fig.~\ref{fig:eta2pippimalp_ee_br}, the corresponding sensitivity curves for $c_{gg}/f_{a}$ and  $c_{qq}/f_{a}$ are shown in Fig.~\ref{fig:eta2pipiALP_cgg_cqq}, for the final states considered in this work.
\end{sloppypar}

\subsubsection{The Heavy-Neutral-Lepton Portal}
The heavy-neutral-lepton portal (HNL) introduces one or more dark fermions that mix with Standard Model neutrinos, offering a natural framework for addressing neutrino masses and BSM phenomena. In the current models, attention is primarily given to realizations within the Two-Higgs Doublet Model (2HDM), which predicts distinctive signatures in rare $\eta$ and $\eta^{\prime}$ meson decays~\cite{Abdallah:2020vgg}. The specific process explored involves decays $\eta/\eta^{\prime} \to \pi^{0} H$, followed by $H \to \nu N_2$ and subsequent transitions $N_2 \to h^{\prime} N_1$ with $h^{\prime} \to e^+e^-$.

\begin{figure}[t]
\centering%
\includegraphics[scale=0.3]{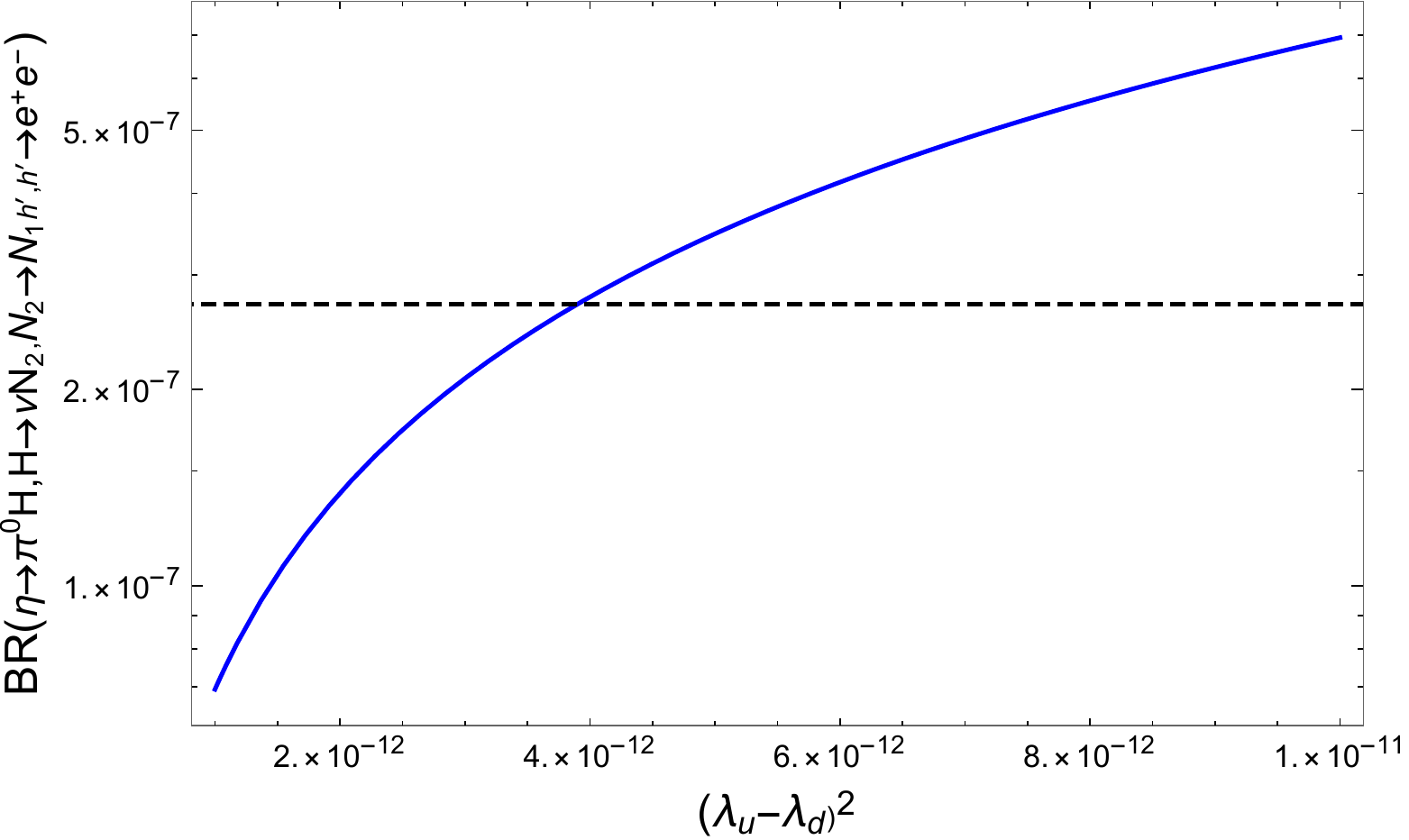} 
\caption{Branching ratio for the process $\eta\rightarrow\pi^{0}H$, $H\rightarrow\nu N_{2}$, $N_{2}\rightarrow N_{1}h'$, $h'\rightarrow e^{+}e^{-}$
predicted by the Two-Higgs-Doublet model~\cite{Abdallah:2020vgg} as a function of $(\lambda_u-\lambda_d)^2$. The dashed line corresponds to the experimental limit for REDTOP with an integrated luminosity of 3.3$\times10^{18}$~POT.}
\label{fig:sensitivity-HNL}
\end{figure}

REDTOP branching ratio sensitivity for the above process is of the order of $\mathcal{O}(10^{-7})$.  
This is somewhat weaker than for the other portals discussed, primarily due to the presence of two missing particles ($\nu$ and $N_1$), which prevents us from constraining the kinematics to the mass of the $\eta$ meson. The sensitivity to $\lambda_u - \lambda_d$ is shown in Fig.~\ref{fig:sensitivity-HNL} superimposed to the prediction of the theoretical model. 
Within the Two-Higgs Doublet Model framework and under the assumption of $\lambda_u = \lambda_d$, the predicted branching ratio for these channels is on the order of $\mathcal{O}(10^{-13})$. But when $\lambda_u \neq \lambda_d$ the branching ratios for $H$ along with those for $N_2$ and $h^{\prime}$, are at the level of 10\(^{-12}\), they are within the reach of the REDTOP experiment. 

\subsection{Tests of Conservation Laws}\label{conservationlaws}
In addition to searches for new particles via hidden portals, precision studies of fundamental symmetries represent a complementary and equally important approach to probing Standard Model limits and physics beyond the Standard Model. Investigations of discrete symmetries, such as $C$, $P$, and $CP$, through rare $\eta$ and $\eta^{\prime}$ decays offer unique sensitivity to potential symmetry violations and provide an essential test bench for many BSM scenarios. Several $\eta$ and $\eta^{\prime}$ decay channels have been identified as crucial for testing fundamental conservation laws. Current studies focus primarily on $CP$ violation, lepton flavor universality, and lepton flavor violation. These categories are supported by well developed theoretical frameworks that are simultaneously consistent with EDM constraints and motivated by unresolved experimental anomalies. The sensitivity studies for the chosen processes and their associated theoretical models are discussed below.

\subsubsection{\label{subsec:CPtests} Tests of \textit{CP} Symmetry} 
Violation of $CP$ symmetry has been extensively studied in flavor-changing decays of neutral $K$- and $B$-mesons~\cite{Hocker:2006xb}. Within the Standard Model, $CP$ violation originates from a single complex phase in the Cabibbo--Kobayashi--Maskawa (CKM) quark mixing matrix. While experimental results in the $K$ and $B$ sectors are largely consistent with CKM predictions, the underlying origin of $CP$ violation remains an open question, and emerging tensions motivate searches for additional sources beyond the Standard Model~\cite{Buras:2009us}. In this context, flavor-conserving processes provide a complementary and theoretically clean way to probe new sources of $CP$ violation, particularly in channels that are not constrained by existing bounds from electric dipole moment (EDM) measurements~\cite{Pospelov:2005pr}.

The REDTOP experiment is designed to explore such flavor-conserving processes with unprecedented sensitivity. Among them, several $\eta$ and $\eta^{\prime}$ decay modes are of particular interest, as they probe distinct classes of $CP$-violating operators that arise from physics beyond the Standard Model. In the following, we focus on the study of charge asymmetric observables in the decay $\eta\to\pi^+\pi^-\pi^0$, which provides a well established and theoretically clean probe of $C$ and $CP$ violation~\cite{Layter:1972aq,Zielinski:2012lsn,Gardner:2019nid,Akdag:2021efj,Akdag:2022sbn,Shi:2024yfa}.
The $\eta\to\pi^+\pi^-\pi^0$ decay proceeds predominantly through strong interactions and is allowed only due to isospin breaking, since Bose symmetry forbids a three-pion final state with $J^{PC}=0^{-+}$ and total isospin $I=0$. As a consequence, the dominant Standard-Model amplitude conserves charge conjugation while violating isospin. A charge asymmetry in the Dalitz plot can arise from the interference between this $C$-conserving, isospin-breaking amplitude and a $C$-violating amplitude. Since parity is conserved in this decay, the observation of a nonzero charge asymmetry constitutes direct evidence for the violation of both $C$ and $CP$ symmetries. Within the Standard Model, contributions to $CP$ violation in this channel from flavor-changing weak interactions are expected to be negligibly small. Therefore, any observable charge asymmetry in $\eta\to\pi^+\pi^-\pi^0$ provides a particularly clean and sensitive probe of BSM physics.

\begin{sloppypar}
Effective field theory (EFT) provides a model independent framework for BSM contributions to flavor conserving $C$ and $CP$ violation. Two complementary approaches have been developed, first which the Standard Model effective field theory (SMEFT)~\cite{Shi:2024yfa}, formulated above the electroweak scale, and the second the low-energy effective field theory (LEFT) below the weak gauge boson masses~\cite{Akdag:2022sbn,Akdag:2023pwx}\footnote{The two analyses differ in the dimension of the SMEFT operators considered as the leading BSM contributions: Ref.~\cite{Shi:2024yfa} uses dimension-six SMEFT, leading to $1/\Lambda^2$ scaling, while Ref.~\cite{Akdag:2022sbn} systematically considers all $C$- and $CP$-violating LEFT operators up to dimension-eight (corresponding to dimension-eight SMEFT), leading to $1/\Lambda^4$ scaling.}. 

In the present analysis, we follow the SMEFT-based parameterization of Ref.~\cite{Shi:2024yfa}, adopting the convention of the original framework of Ref.~\cite{Gardner:2019nid} for the assignment of the free parameters therein. In this convention, the $C$- and $CP$-violating contributions to the Dalitz plot distribution at lowest order in Chiral Perturbation Theory ($\chi$PT) are parameterized by two real low-energy coefficients, $\alpha$ and $\beta$, associated with the $I=0$ (appearing at $\mathcal{O}(p^6)$) and $I=2$ (appearing at $\mathcal{O}(p^2)$) amplitudes, respectively. These coefficients encode the dependence on the new physics scale $\Lambda$ as $1/\Lambda^2$, so that their measurement directly probes the BSM scale of $C$ and $CP$ violation in flavor-conserving processes.
The lowest order $I=0$ amplitude appears only at $\mathcal{O}(p^6)$ in $\chi$PT, while the $I=2$ amplitude already contributes at $\mathcal{O}(p^2)$.
As a consequence, even a relatively small $\beta$ produces a sizable imprint on the Dalitz plot, whereas a large $\alpha$ may yield only a marginal effect: the experimental sensitivity to $\alpha$ is therefore intrinsically weaker than to $\beta$. 

Crucially, the observable charge asymmetry is linear in these $CP$-violating parameters, rather than quadratic as in branching ratio measurements of pure $CP$-violating processes, making this channel particularly sensitive to small BSM effects.
Complementary indirect constraints on the $I=0$ sector can be obtained from the semileptonic decay $\eta \to \pi^0 \ell^+ \ell^-$, providing bounds of comparable strength to those from $\eta \to \pi^+\pi^-\pi^0$~\cite{Akdag:2023pwx}.

To evaluate the achievable precision on these parameters at REDTOP, simulated $\eta\to\pi^+\pi^-\pi^0$ data samples were generated and fitted under various experimental conditions, following the procedure of Ref.~\cite{Shi:2024yfa}. 
The simulation procedure follows the SMEFT-based framework which employs an NLO $\chi$PT description of the $C$-conserving Standard-Model amplitude~\cite{Gasser:1984pr}. A more refined dispersive treatment, which resums the final-state $\pi\pi$ rescattering effects to all orders, is available in Ref.~\cite{Akdag:2022sbn} and will be incorporated in future analyses to match the precision required by high-statistics REDTOP measurements.

The generated samples assume the Standard-Model expectation, $\alpha=\beta=0$, so that the resulting fit uncertainties directly reflect the experimental sensitivity to $CP$ violation. The statistical uncertainties on $\alpha$ and $\beta$ are summarized in Table~\ref{table:recoeff_eta3pi-fit-SMEFT} for a baseline exposure of $3.3\times10^{8}$ protons on target (POT), as well as for full statistics scenarios with and without background contamination. 
Extrapolating to the full integrated luminosity expected at REDTOP, corresponding to approximately $3.3\times10^{18}$ POT and about $2.3\times10^{13}$ reconstructed $\eta\to\pi^+\pi^-\pi^0$ decays, the statistical uncertainties on the $CP$-violating parameters are projected to reach the level of $\sigma(\alpha) \sim \mathcal{O}(1)~\text{GeV}^{-6}$ and $\sigma(\beta) \sim \mathcal{O}(10^{-3})~\text{GeV}^{-2}$. 
This represents an improvement of more than an order of magnitude compared to existing experimental constraints from KLOE-2~\cite{KLOE-2:2016zfv} and BESIII~\cite{BESIII:2023edk}, and would either reveal a nonzero $CP$-violating signal in $\eta$ decays or establish significantly more stringent upper limits on BSM $CP$ violation in this channel.
\begin{table}[t]
\centering
\renewcommand{\arraystretch}{1.2}
\begin{tabular}{lcc}
\toprule
Reconstructed events &
$\sigma(\alpha)$[GeV$^{-6}$] &
$\sigma(\beta)$[GeV$^{-2}$] \\
\midrule
$10^8$ (no-bkg) &
$1.9\times10^{1}$ &
$1.5\times10^{-2}$ \\
Full stat. (no-bkg) &
$1.1$ &
$8.3\times10^{-4}$ \\
Full stat. (100\% bkg) &
$1.4$ &
$1.2\times10^{-3}$ \\
\bottomrule
\end{tabular}
\caption{Projected statistical sensitivities of the REDTOP experiment to the $C$- and $CP$-violating parameters $\alpha$ (associated with the $I=0$ amplitude at $\mathcal{O}(p^6)$) and $\beta$ (associated with the $I=2$ amplitude at $\mathcal{O}(p^2)$) in $\eta\to\pi^+\pi^-\pi^0$ decay, in the SMEFT-based framework of Ref.~\cite{Shi:2024yfa}. The simulated samples assume the Standard Model expectation $\alpha=\beta=0$. }
\label{table:recoeff_eta3pi-fit-SMEFT}
\end{table}

Based on the current experimental constraints from previous KLOE-2~\cite{KLOE-2:2016zfv} and new BESIII~\cite{BESIII:2023edk} data, the accessible new physics scale is estimated to be $\Lambda \gtrsim \mathcal{O}(1~\text{GeV})$, which is significantly below the TeV scale expected for BSM physics. This reflects the limited statistical power of existing data rather than a fundamental limitation of the method. The situation changes significantly with the statistics anticipated at REDTOP. As shown in Ref.~\cite{Shi:2024yfa}, probing new physics at $\Lambda \sim 1~\text{TeV}$ requires experimental samples of at least $\mathcal{O}(10^{13})$ $\eta\to\pi^+\pi^-\pi^0$ decays. With this projected statistics, REDTOP is therefore positioned at the threshold of TeV-scale .  

As already noted, the flavor-conserving nature of $\eta\to\pi^+\pi^-\pi^0$ decay makes this channel complementary to other probes of BSM $CP$ violation, such as electric dipole moment measurements, which are sensitive to $P$- and $T$-violating operators. Since the observable charge asymmetry is linear in the $CP$-violating coefficients, REDTOP benefits from enhanced sensitivity to small BSM effects in flavor-conserving $C$ and $CP$ violation searches.
\end{sloppypar}

\subsubsection{\label{subsec:LFU} Lepton Flavor Universality} 
\begin{sloppypar}
Lepton Flavor Universality (LFU), the equality of gauge couplings of electrons, muons, and tau leptons, is a fundamental prediction of the Standard Model. Sensitive tests of LFU can be performed by comparing $\eta^{(\prime)}$ decays into different charged lepton flavors, e.g., the ratio
\begin{equation}
R_{\mu/e}^{X} = \frac{\Gamma\!\left(\eta^{(\prime)} \to X\,\mu^+\mu^-\right)}{\Gamma\!\left(\eta^{(\prime)} \to X\,e^+e^-\right)},
\end{equation}
where $X$ may be $\gamma$, $\pi^0$, $\eta$, or another lepton pair. 
In the Standard Model, weak-interaction-induced decays of this type are extremely rare, with branching ratios below $10^{-10}$, several orders of magnitude smaller than the current experimental limits~\cite{Gan:2020aco}.  
For purely leptonic modes, the dominant SM contribution proceeds through intermediate two-photon exchange and is therefore strongly suppressed.\footnote{This statement, in fact, also holds true for the loop-induced SM mechanism for the semileptonic decays $\eta^{(\prime)}\to\pi^0\ell^+\ell^-$, $\eta'\to\eta\ell^+\ell^-$~\cite{Escribano:2020rfs,Schafer:2023qtl}.}  
This suppression makes these decays particularly sensitive to possible BSM contributions in the MeV--GeV mass range. From the experimental perspective, both leptonic and semileptonic $\eta$ decays provide clean signatures at REDTOP.  
The detector high granularity and excellent particle identification capabilities allow for the efficient separation of such channels from the large hadronic background.  
Moreover, the partial widths into electrons and muons differ only slightly due to phase space effects, enabling precise LFU tests through their ratios. REDTOP will explore LFU by measuring two complementary classes of processes:
$\eta \to  \ell_1^+ \ell_1^- \ell_2^+ \ell_2^-$, and 
$\eta \to  \gamma \ell^+ \ell^-$.
which together provide sensitivity to both interactions potentially violating lepton universality.

The $\eta$ decays into four charged leptons have all been observed experimentally: $\eta \to e^{+}e^{-}e^{+}e^{-}$ by KLOE, with a measured branching ratio of ${\rm BR} = (2.4 \pm 0.2_\text{stat} \pm 0.1_\text{syst})\times10^{-5}$~\cite{KLOE:2011qwm}, 
and BESIII, with ${\rm BR} = (2.63 \pm 0.34_\text{stat} \pm 0.16_\text{syst})\times10^{-5}$~\cite{BESIII:2026izr};
the mixed channel $\eta\to e^+e^-\mu^+\mu^-$ by CMS, with ${\rm BR} = (2.4 \pm 0.3_\text{stat} \pm 0.6_\text{syst} \pm 0.3_{\text{BR}(2\mu\gamma)})\times10^{-6}$~\cite{CMS:2026qbd} (where the last uncertainty denotes the one in the normalization channel $\eta\to\gamma\mu^+\mu^-$); and the rare four-muon decay $\eta\to\mu^{+}\mu^{-}\mu^{+}\mu^{-}$, also observed by CMS with ${\rm BR} = (5.0 \pm 0.8_\text{stat} \pm 0.7_\text{syst})\times10^{-9}$~\cite{CMS:2023thf}. These measurements are consistent with SM predictions, providing a baseline against which BSM contributions can be probed. 
Given the projected REDTOP dataset, this corresponds to an expected sample of roughly $2.6\times10^{9}$ reconstructed events, providing an opportunity to study this decay with high precision and to search for possible contributions from LFU or non-Standard-Model interactions. Comparable estimates for the other four-lepton final states have been obtained in Ref.~\cite{Kampf:2018wau}. For the mixed channel $\eta \to e^{+}e^{-}\mu^{+}\mu^{-}$, the expected yield at REDTOP is on the order of $\mathcal{O}(10^{7})$, while for the fully muonic mode $\eta \to \mu^{+}\mu^{-}\mu^{+}\mu^{-}$, it is approximately $\mathcal{O}(10^{4})$. These projections were derived from dedicated simulations of $\eta$ meson production under REDTOP running conditions, with the corresponding POT statistics summarized in the second column of Table~\ref{table:staterrot_eta4lepton}.
\begin{table}[t]
\centering
\renewcommand{\arraystretch}{1.2}
\begin{tabular}{ccc|c}
\toprule 
Process & POT & Signal & Statistical\\
 &  & events & error\\
\midrule 
$\eta\rightarrow e^{+}e^{-}e^{+}e^{-}$ &  $4.4\times10^{14}$ & 53,934 & 0.5\%\\ 
$\eta\rightarrow e^{+}e^{-}\mu^{+}\mu^{-}$ &  $1.6\times10^{15}$  & 18,841 & 0.8\%\\ 
 $\eta\rightarrow\mu^{+}\mu^{-}\mu^{+}\mu^{-}$ &  $2.2\times10^{18}$  & 10,548 & 1.0\%\\ 
 \bottomrule
\end{tabular}
\caption{Statistical error from the fit of $\eta\rightarrow \ell_1^+ \ell_1^- \ell_2^+ \ell_2^-$ and UrQMD generated background using a gaussian and a 5th-order polynomial. The POT corresponding to each data sample is indicated in the second column.}
\label{table:staterrot_eta4lepton}
\end{table}

When the full event statistics are taken into account, the projected statistical uncertainty for the two channels involving electrons in the final state is expected to reach the $10^{-5}$ level. At this precision, the measurement becomes limited primarily by systematic effects such as acceptance corrections, lepton identification efficiencies, and radiative modeling. Such accuracy would enable REDTOP to probe potential deviations in the electron--muon universality ratio down to the few-per-mille level, comparable to or exceeding the sensitivities achieved in kaon and pion decays.  In addition, the large available statistics open the possibility of differential studies of the dilepton invariant-mass distributions, which are particularly sensitive to loop-induced or contact-type new physics operators.

Another radiative dileptonic decay providing a particularly clean probe of LFU is $\eta \to \gamma \ell^+ \ell^-$.
The presence of an additional photon introduces new Standard-Model contributions, however, the main conclusions drawn from the purely leptonic case remain valid. A key difference is that the chiral suppression inherent to nonradiative decays~\cite{Jarlskog:1967fpu,Masjuan:2015cjl,Messerli:2025rnv} is lifted through the emission of the extra photon. Consequently, ratios of radiative dileptonic decay rates, where the numerator and denominator differ only by the lepton flavor, provide excellent tests of Lepton Flavor Universality (LFU)~\cite{Guadagnoli:2016erb}. Within the Standard Model, these ratios are expected to be extremely close to unity. Any observed deviation from unity in these ratios would, therefore, indicate nonuniversal couplings of new particles to charged leptons.

To evaluate the experimental sensitivity of these radiative channels, dedicated simulations of $\eta \to \gamma\, \ell^+\ell^-$ decays were performed.  
The studies were conducted using a representative sample of $4.65\times10^{6}$ generated $\eta$ mesons, corresponding to approximately $4\times10^{-8}$ of the total integrated luminosity anticipated for REDTOP. Statistical uncertainties on the branching ratios were obtained from fits to the invariant mass distributions of the $\eta \to \gamma\, \ell^+\ell^-$ system for each lepton flavor channel. The resulting uncertainties are summarized in Table~\ref{table:staterrot_etag2lepton}.
\begin{table}[t]
\centering
\renewcommand{\arraystretch}{1.2}
\begin{tabular}{cccc|c}
\toprule 
{Process} & {Signal} & {Back.} & {${S}/{\sqrt{B}}$} & Stat.\\
 & events & events &  & error\\
\midrule 
{$\eta\rightarrow\gamma\,e^{+}e^{-}$} & {$2.13\times10^{6}$} & {$2.52\times10^{4}$} & {$1.3\times10^{4}$} & {0.09\%}\\
\hline 
{$\eta\rightarrow\gamma\,\mu^{+}\mu^{-}$} & {$8.84\times10^{5}$} & {$6.50\times10^{3}$} & {$3.5\times10^{3}$} & {0.14\%}\\
\bottomrule 
\end{tabular}
\caption{Statistical error from the fit of $\eta\rightarrow\gamma \ell^+ \ell^-$ and UrQMD generated background using a gaussian and a 5th-order polynomial, for 1.38$\times10^{18}$ POT.}
\label{table:staterrot_etag2lepton}
\end{table}
When the full REDTOP statistics of $3.3\times10^{18}$ POT are taken into account, the projected statistical uncertainty for the $\eta\!\to\!\gamma\,\ell^+\ell^-$ final state is expected to reach the $10^{-6}$ level thus, it becomes negligible compared to systematic effects. At this precision, the measurement will be limited primarily by detector related and radiative correction uncertainties. Such accuracy will enable highly sensitive tests of lepton flavor universality and provide an opportunity to search for small deviations in the $\eta$ electromagnetic transition form factor~\cite{Masjuan:2017tvw,Eichmann:2019tjk,Leutgeb:2022lqw,Estrada:2024cfy,Holz:2024lom,Holz:2024diw} that could signal contributions from new light states.
\end{sloppypar}

\subsection{Muon Polarimetry\label{sec:muon-polarimetry}}
Beyond rare decay measurements, the ability to measure the polarization of muons produced in $\eta$ and $\eta^{\prime}$ decays offers an additional and powerful probe of discrete symmetry violation and possible new interactions beyond the Standard Model. A striking consequence of the quantum numbers of the $\eta$/$\eta^{\prime}$ mesons is that, for selected decays, certain components of the muon polarization, such as longitudinal or transversal, are highly suppressed within the Standard Model. This opens the door to probe a broad class of new physics scenarios, which could manifest as nonzero polarization of muons originating from $\eta$/$\eta^{\prime}$ decays. In particular, polarization observables in processes such as 
$\eta \to \mu^{+}\mu^{-} (\gamma/\pi^0)$ and $\eta \to \pi^{+}\pi^{-}\mu^{+}\mu^{-}$ are sensitive to interference terms between parity conserving and parity violating amplitudes and can therefore reveal small $CP$- or $T$-odd contributions that may not be accessible through rate measurements alone. Muon polarimetry provides a unique handle on these effects, since the decay $\mu^{+}\!\to\!e^{+}\nu_{e}\bar{\nu}_{\mu}$ preserves the spin information of the parent muon through the angular distribution of the emitted positron. By reconstructing this distribution with sufficient precision, REDTOP can directly measure the polarization of muons produced in $\eta$ decays.  

\subsubsection{\textit{CP} Violation in \texorpdfstring{$P_T$}{PT} or \texorpdfstring{$P_L$}{PL} Muon Polarization}
\begin{sloppypar}
In the decay $\eta \to \mu^+ \mu^- X$, a muon pair can be produced in different angular momentum configurations, depending on the underlying production mechanism. 
In the Standard Model, the radiative decay ($X = \gamma$) is dominated by virtual-photon exchange ($\eta \to \gamma \gamma^* \to \gamma \mu^+\mu^-$), 
so that the $\mu^+\mu^-$ pair inherits the photon's quantum numbers $J^{PC} = 1^{--}$ and is produced in a $^3S_1$ state. In the semileptonic decay ($X = \pi^0$), in contrast, cannot proceed through a single virtual-photon exchange, as $C$ conservation forbids the $\eta \to \pi^0 \gamma^*$ transition (cf., e.g., Ref.~\cite{Akdag:2023pwx}). In the SM it instead proceeds through two-photon exchange $\eta \to \pi^0 \gamma^* \gamma^* \to \pi^0 \mu^+\mu^-$, which is suppressed~\cite{Escribano:2020rfs,Schafer:2023qtl}. In contrast, the purely leptonic decay ($X = \emptyset$) requires the muon pair to carry the quantum numbers of the parent meson, $J^{PC} = 0^{-+}$, corresponding to a $^1S_0$ state. 
Beyond the Standard Model, additional configurations become accessible where a (pseudo)scalar quark--lepton operator can produce the muon pair in either a $^1S_0$ ($CP$-conserving) or $^3P_0$ ($CP$-violating) state, with the precise assignment of $CP$ properties depending on the quantum numbers of $X$ and the orbital angular momentum between the muon pair and $X$. In what follows, we focus on the BSM scenario in which the muon pair is produced through (pseudo)scalar quark--lepton operators~\cite{Sanchez-Puertas:2018tnp}, and we use the resulting muon polarization observables to assess the $CP$-violating sensitivity at REDTOP.
\end{sloppypar}

\begin{sloppypar}
For BSM contributions where the muon pair couples through (pseudo) scalar quark-lepton operators, the amplitude can be expressed as~\cite{Geng:1989zk,Geng:1990dw,Sanchez-Puertas:2018tnp}:
\begin{equation}
\mathcal{M} = g_P\,(\bar{u}\,i\gamma^5 v) + g_S\,(\bar{u}\,v),
\end{equation}
where $g_P$ and $g_S$ are dimensionless couplings parameterizing the pseudoscalar and scalar muon pair structures, respectively. Both terms are $C$-even but have opposite $P$ transformations, so their interference generates a $CP$-violation observable in the muon polarization. This parametrization is directly applicable to $\eta \to \mu^+\mu^-$, where the muon pair must carry the quantum numbers of the parent meson. For the radiative $\eta \to \gamma \mu^+\mu^-$ and semileptonic $\eta \to \pi^0 \mu^+\mu^-$ channels, the dominant Standard Model contribution proceeds through virtual-photon exchange, in which the $\mu^+\mu^-$ pair is produced in a $^3S_1$ state, and a different formalism is required for the corresponding $CP$-conserving amplitude.
\end{sloppypar}

\begin{sloppypar}
The polarized differential decay width for $\eta \to \mu^+\mu^-$ can be expressed as~\cite{Sanchez-Puertas:2018tnp,Masjuan:2015cjl}:
\begin{align}
\nonumber    d\Gamma &= \frac{\beta_{\mu}}{16\pi m_{\eta}} \times 
    \frac{m_{\eta}^2}{2}\Big[
    |g_P|^2(1-s^+\!\cdot s^-)\\
\nonumber     &+|g_S|^2\beta_{\mu}^2\left[1 - \left\{2\big(s^+\!\cdot \hat{\boldsymbol{\beta}}_{\mu^+}\big)\big(s^-\!\cdot \hat{\boldsymbol{\beta}}_{\mu^+}\big) - s^+ \! \cdot s^- \right\}\right] \\
    &+2\operatorname{Re}(g_Pg_S^*)\boldsymbol{\beta}_{\mu^+}\!\cdot(s^-\!\times s^+) \notag\\
    &+2\operatorname{Im}(g_Pg_S^*)\boldsymbol{\beta}_{\mu^+}\!\cdot(s^+ -s^-)
    \Big],
\end{align}
where $\boldsymbol{\beta}_{\mu^+}$ is the velocity of the $\mu^+$, with magnitude $\beta_{\mu} = \sqrt{1 - 4m_{\mu}^2/m_{\eta}^2}$, and $s^{\pm}$ denote the spin vectors of the $\mu^{\pm}$. The first two terms describe $CP$-conserving contributions, while the latter two originate from the interference between scalar and pseudoscalar couplings and thus carry $CP$-odd or $T$-odd correlations.
\end{sloppypar}

\begin{sloppypar}
Experimentally, these effects can be quantified through the longitudinal $P_{L}$ and transverse $P_T$ components of the muon polarization, defined respectively as:
\begin{equation}\label{eq:polL}
P_{L} = \frac{N_R - N_L}{N_R + N_L}, \qquad P_T = \frac{N_{RH} - N_{LH}}{N_{RH} + N_{LH}},
\end{equation}
where $N_{R(L)}$ denotes the number of $\mu^+$ with positive (negative) helicity, and $N_{RH(LH)} = (s^+ \times s^-)\!\cdot\!\boldsymbol{\beta}_{\mu^+}$ represents the number with positive (negative) transverse polarization. Notably, $P_L$ is $C$-even, $P$-odd, and $T$-even, while $P_T$ is $C$-even, $P$-odd, and $T$-odd. Thus, a nonzero value of either observable would provide an unambiguous indication of $CP$ violation originating from interference between scalar and pseudoscalar amplitudes in $\eta$ decays. Within the Standard Model, such effects are expected to be vanishingly small, since $CP$ violation arises solely through the complex phase of the CKM matrix and is further suppressed by weak interaction factors. Consequently, any observation of a nonzero polarization would unambiguously signal the presence of physics beyond the Standard Model.  
\end{sloppypar}

To quantify the achievable experimental sensitivity for REDTOP, three decay channels have been investigated using Monte Carlo simulations:  
$\eta\!\to\!\mu^+\mu^-$, 
$\eta\!\to\!\pi^{0}\mu^{+}\mu^{-}$, and 
$\eta\!\to\!\gamma\mu^{+}\mu^{-}$. 
In the performed simulation studies, a conservative muon polarization reconstruction efficiency of $\epsilon_{\mathrm{pol}} = 50\%$ was assumed for both signal and background events. The $\eta\!\to\!\mu^+\mu^-$ channel, discussed below, serves as a benchmark case for assessing the achievable sensitivity to $CP$-violating effects.  
\begin{figure}[t]     
\centering     
\includegraphics[width=0.42\textwidth]{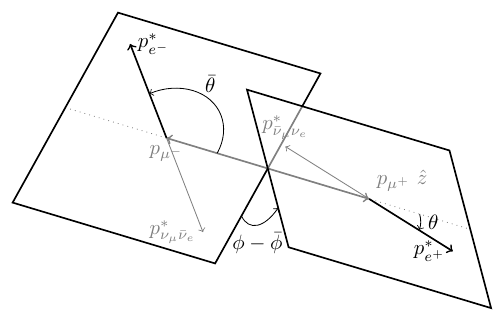}     
\caption{Kinematics of the $\eta\!\to\!\mu^+\mu^-$ process. The decaying muons momenta in the $\eta$ rest frame are noted as $p_{\mu^{\pm}}$, while the $e^{\pm}$ momenta, $p^*_{e^{\pm}}$, is shown in the corresponding $\mu^{\pm}$ reference frame along with the momenta of the $\nu\bar{\nu}$ system. The $\hat{z}$ axis is chosen along $p_{\mu^+}$.}    
\label{fig:EtaTo2L}
\end{figure} 

Two polarization asymmetries of the muon that can be probed experimentally may be connected to a SMEFT expansion of the Lagrangian of the underlying physical process. They correspond to longitudinal and transverse polarizations~\cite{Sanchez-Puertas:2018tnp}:
\begin{align}     
\nonumber A_{L} &= \frac{N(\cos\theta >0) -N(\cos\theta <0)}{N} \\&= \operatorname{Im}\left[4.1\cdot c_{\ell edq}^{2222} - 2.7\cdot\left(c_{\ell equ}^{2211} +c_{\ell edq}^{2211}\right)\right]\times10^{-2}, \label{eq:AL}\\      
\nonumber A_{\times} &= \frac{N(\sin\Phi >0) -N(\sin\Phi <0)}{N} \\ &=\operatorname{Im}\left[2.5\cdot c_{\ell edq}^{2222}-1.6\cdot \left(c_{\ell equ}^{(1)2211} +c_{\ell edq}^{2211}\right)\right]\times10^{-3}, \label{eq:AT}
\end{align}
where, $\theta$ is the angle between the $e^{+}$ and the $\mu^{+}$ momenta in the muon rest frame, with the z-axis defined in the $\eta$ rest frame. Angle $\Phi = \phi - \bar{\phi}$ encodes the $CP$-violating observable via the sign of $(\vec{p}_{e^-} \times \vec{p}_{e^+}) \cdot \vec{p}_{\mu^+}$, and the  $c_{\mathcal{O}}^{(llqq)}$ are Wilson coefficients for the associated SMEFT operators. 
The asymmetries provide sensitivity to those that violate $CP$ without being constrained by EDM bounds, particularly in the muon sector. To measure these asymmetries, the reconstruction of the direction of the electron from the muon decay $\mu^\pm \to e^\pm \nu \bar{\nu}$ is needed. The numerical coefficients in Eqs.~\eqref{eq:AL}--\eqref{eq:AT} correspond to the $\eta \to \mu^+\mu^-$ channel, while analogous expressions for the radiative ($\eta \to \gamma \mu^+\mu^-$) and semileptonic ($\eta \to \pi^0 \mu^+\mu^-$) channels, which are dominated by virtual-photon exchange in the Standard Model, are obtained using the corresponding SMEFT operator matching as discussed in Refs.~\cite{Sanchez-Puertas:2018tnp,Escribano:2024odh}.

\begin{sloppypar}
Combining the results and assuming $\epsilon_{\text{pol}} = 50\%$, the statistical uncertainty of $A_L$ is:
\begin{equation}
\sigma (A_L) = \frac{\sqrt{N_{\text{sig}}} + \sqrt{N_{\text{bkg}}}}{N_{\text{sig}}} = 2.7 \times 10^{-3}.
\label{eq:delta-a-l}
\end{equation}
Comparing the obtained values to Eq.~\eqref{eq:AL}, we find the REDTOP expected sensitivity to the Wilson coefficients for the three processes under consideration:\\
$\eta\!\to\!\mu^+\mu^-$
\begin{equation}
\nonumber
\Delta(c_{\ell equ}^{1122})=0.01,~\Delta(c_{\ell edq}^{1122})=0.1,~\Delta(c_{\ell edq}^{2222})=6.6\times10^{-2}\label{eq:error-on-c2222},
\end{equation}
$\eta\!\to\!\gamma\mu^{+}\mu^{-}$:
\begin{equation}
\nonumber
\Delta(c_{\ell equ}^{1122})=2.6,\quad\Delta(c_{\ell edq}^{1122})=2.6,\quad\Delta(c_{\ell edq}^{2222})=1.7\label{eq:error-on-c2222_a},
\end{equation}
$\eta\!\to\!\pi^{0}\mu^{+}\mu^{-}$:
\begin{equation}
\nonumber
\Delta(c_{\ell equ}^{1122})=21,\quad\Delta(c_{\ell edq}^{1122})=21,\quad\Delta(c_{\ell edq}^{2222})=200\times10^{-2}\label{eq:error-on-c2222_b}.
\end{equation}
The result for $c_{\ell edq}^{2222}$ from the $\eta\!\to\!\mu^+\mu^-$ process, which is responsible for $CP$ violation from sources outside the SM, is competitive with constraints from neutron EDM bounds~\cite{Abel:2020gbr}, affirming the good precision potential of REDTOP $CP$ violation studies in $\eta \to \mu^+\mu^-$. A more detailed discussion on this topic can be found in Ref.~\cite{Escribano:2024odh}.
\end{sloppypar}

\subsubsection{\textit{T} Violation in \texorpdfstring{$P_T$}{PT} Muon Polarization}
\begin{figure}[tb]
\centering{}
\includegraphics[width=0.49\columnwidth]{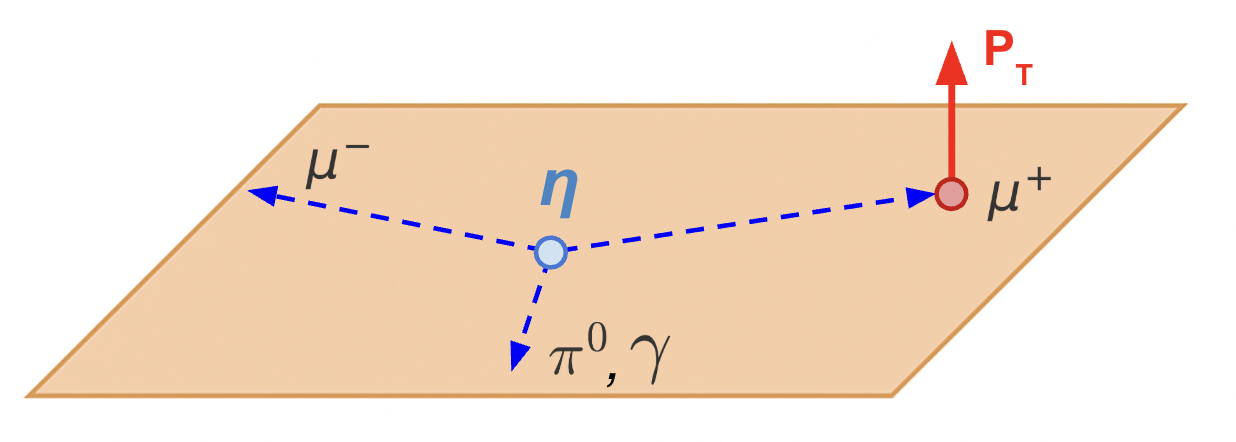}
\includegraphics[width=0.49\columnwidth]{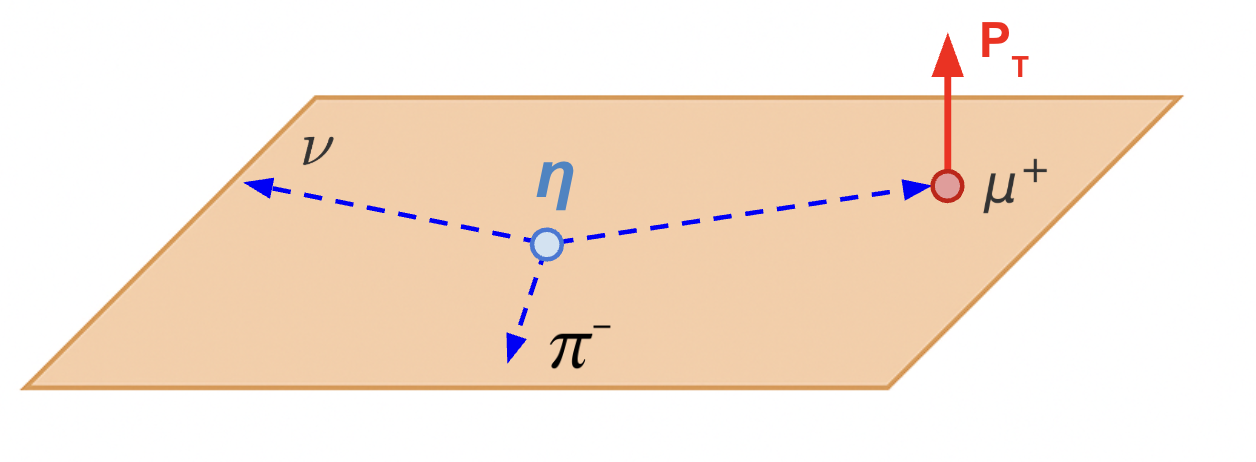}
\caption{
Transverse muon polarization $\vec{P}_T$ in $\eta$ decays 
at rest: (left) semileptonic ($\eta \to \pi^0 \mu^+ \mu^-$) and radiative 
($\eta \to \gamma \mu^+ \mu^-$) channels, (right) semileptonic 
$\eta \to \pi^- \mu^+ \nu$ channel. $\vec{P}_T$ is the component 
of the $\mu^+$ spin polarization perpendicular to the decay plane 
defined by the three-momenta of the final-state particles. }
\label{fig:eta2muon-pt} 
\end{figure}

\begin{sloppypar} 
As noted in the previous section, the transverse muon polarization $P_T$ in meson decays is a $T$-odd observable, defined as the projection of the muon spin perpendicular to the decay plane. A nonzero value of $P_T$ would be a clear signature of time reversal ($T$) violation~\cite{Sakharov:1967dj}, as spurious contributions from final-state interactions are known to be small~\cite{Zhitnitsky:1980tq}. Time reversal violation is of particular interest because, under the $CPT$ theorem, it implies $CP$ violation as well. Such $CP$ violating sources beyond the SM are among the Sakharov conditions~\cite{Sakharov:1967dj} for explaining the observed baryon asymmetry of the Universe. While the SM accommodates $CP$ violation via the CKM matrix, this source is insufficient to explain the observed asymmetry. Therefore, measuring $T$-odd observables such as $P_T$ is a promising path toward discovering new physics. The transverse muon polarization in $\eta/\eta^{\prime}$ decays is illustrated in Fig.~\ref{fig:eta2muon-pt} for both Dalitz (left) and semileptonic (right) decays.
\end{sloppypar}

In the SM, predictions for $P_T$ are extremely small, typically arising only at higher orders. In the kaon decays, $P_T$ has been computed to be at the level of $10^{-5}$~\cite{Zhitnitsky:1980tq,Efrosinin:2000yv}, and a similar suppression is expected in $\eta \to \pi^- \mu^+ \nu_\mu$ decays~\cite{Bigi1999}. In contrast, models involving new physics—such as multi-Higgs doublet models, leptoquark scenarios, or supersymmetry with R-parity violation—can predict significantly larger values of $P_T$, ranging from $10^{-4}$ to $10^{-2}$~\cite{Garisto:1991fc,Wu:1996zq}. It is worth noting that the semileptonic decay $\eta \to \pi \mu \nu_\mu$ has not yet been observed.

In the case of Dalitz decays such as $\eta \to \gamma \ell^+ \ell^-$, leading-order QED and weak processes do not induce transverse polarization~\cite{Sanchez-Puertas:2018tnp}, meaning that any measurable $P_T$ would be a clear signal of BSM physics. Similarly, in the decay $\eta \to \pi^0 \mu^+ \mu^-$, no contribution to $P_T$ is expected within the SM~\cite{Escribano:2022zgm}. Thus, any observation of transverse muon polarization in three-body decays of $\eta$ or $\eta^{\prime}$ mesons would constitute unambiguous evidence for $T$-violation, and by extension, $CP$-violation and New Physics.

\subsection{Nonperturbative QCD}
Beyond their potential for probing new physics, the $\eta$ and $\eta^{\prime}$ mesons provide an essential testing ground for the Standard Model itself, particularly in the domain of low-energy, nonperturbative Quantum Chromodynamics (QCD). Decays of the $\eta$ and $\eta^{\prime}$ mesons offer unique access to the dynamics of chiral symmetry breaking and anomaly driven processes, thereby complementing the searches for physics beyond the Standard Model discussed in the preceding sections. For comprehensive reviews on this subject, see, e.g., Refs.~\cite{Gan:2020aco, JPAC:2021rxu}. At energy scales comparable to the $\eta$ and $\eta^{\prime}$ masses (below $\sim$1 GeV), the strong interaction enters a nonperturbative regime in which standard perturbation theory based on an expansion in the strong coupling constant $\alpha_s$ becomes unreliable. In this domain, theoretical investigations rely on a variety of nonperturbative methods, including effective field theories such as $\chi$PT for light quarks, dispersion relation techniques that exploit analyticity and unitarity, and numerical simulations based on lattice QCD. Together, these approaches provide complementary insights into the structure of the $\eta$ and $\eta^{\prime}$ mesons, as well as the dynamics governing their decays.

The $\eta$ meson, in particular, stands out because its dominant decay mode, $\eta \!\to\! 3\pi$, violates $G$-parity—a symmetry combining charge conjugation and isospin rotation. Although this decay proceeds via strong interactions, it is entirely driven by isospin breaking, primarily due to the mass difference between the up and down quarks~\cite{Baur:1995gc,Ditsche:2008cq}. As a result, it provides a sensitive probe of light quark mass ratios and a stringent test of $\chi$PT. In particular, precise measurements of the Dalitz plot parameters and the total decay width can refine the determination of the quark mass ratio~\cite{Bijnens:2007pr,Kampf:2011wr,Guo:2016wsi,Colangelo:2016jmc,Albaladejo:2017hhj,Colangelo:2018jxw}
\begin{equation}
Q^2 = \frac{m_s^2 - \hat{m}^2}{m_d^2 - m_u^2}, 
\quad 
\text{where } \hat{m} = \tfrac{1}{2}(m_u + m_d),
\end{equation}
which plays a central role in quantifying isospin breaking in QCD.
High statistics samples are therefore essential to improve the precision of these measurements and to constrain theoretical predictions at the percent level.

\begin{sloppypar}
The case of the $\eta^{\prime}$ meson is considerably more intricate. Unlike the $\eta$, it cannot be adequately described within the framework of standard $SU(3)$ Chiral Perturbation Theory, owing to its strong coupling to gluonic degrees of freedom through the axial $U(1)_A$ anomaly~\cite{tHooft:1976rip,Witten:1979vv}.  
This anomaly contributes significantly to the $\eta^{\prime}$ mass~\cite{Veneziano:1979ec}, lifting it far above the pseudo-Goldstone boson scale expected from spontaneous chiral symmetry breaking. To incorporate the $\eta^{\prime}$ into a chiral effective framework, extensions such as $U(3)$ $\chi$PT or large-$N_c$ expansions are required, where the anomaly is suppressed in the limit of a large number of colors~\cite{Feldmann:1999uf,Kaiser:2000gs}. However, these approaches remain under active development and introduce additional low-energy constants that are not yet well constrained by data. Many $\eta^{\prime}$ decay modes involve multibody hadronic final states or rare electromagnetic transitions, posing both theoretical and experimental challenges. Such processes are particularly sensitive to gluon dynamics, possible gluonium admixtures, and the interplay between quark and gluon degrees of freedom in hadron structure.
\end{sloppypar}

High-precision measurements from REDTOP, with its large anticipated samples of $\eta^{\prime}$ decays, will provide critical input for testing these theoretical frameworks. The data will constrain the poorly known parameters in extended chiral Lagrangians and help quantify the impact of anomaly related effects, making REDTOP a key experiment for exploring the nonperturbative frontier of QCD. With its projected yields of about $10^{14}$ $\eta$ and $10^{12}$ $\eta^{\prime}$ mesons,  REDTOP will deliver an unprecedented dataset. This will not only enable precision studies of rare decays and discrete symmetry violations, but also supply vital information for refining theoretical tools in nonperturbative QCD, thereby extending the precision frontier of hadronic physics.

\section{Beam Options for REDTOP}\label{sec:beam}
The most efficient production method of $\eta/\eta^{\prime}$ mesons is via the interaction of a high-intensity proton or pion beam with a nuclear target.
Above the production threshold for $\eta/\eta^{\prime}$ mesons ($E_{\text{kin}} \approx 1.3$ GeV), several intra-nuclear baryonic resonances are excited, such as $\Delta$ resonances, $N(1440)$, and $N(1535)$. Some of these resonances decay into an $\eta$ or $\eta^{\prime}$ meson, providing an efficient mechanism for meson production in a fixed-target setup.

The choice of projectile plays a significant role in production efficiency. 
Proton beams tend to favor the excitation of the $N(1535)$ resonance, which has a strong coupling to the $\eta N$ final state, making it particularly effective for $\eta$ production at the expense of a higher QCD background. On the other hand, pion beams, especially $\pi^{-}$, can exploit isospin selectivity and different resonance excitation modes, potentially leading to higher cross sections for $\eta^{\prime}$ in some energy ranges. The QCD background produced by a pion beam is typically lower, with fewer particles in the final state. Consequently, the signal-to-background ratio is expected to be smaller, and the sensitivity to New Physics is higher.

Similarly, the choice of target nucleus affects both the overall yield and the background. Light nuclei, such as hydrogen and deuterium, offer a cleaner environment with better kinematic resolution, allowing for precise reconstruction of the meson decay products. Heavier nuclear targets (e.g., carbon, lithium, or beryllium), while introducing more background and Fermi motion effects, can enhance the total production cross section due to their larger nucleon content and multiple scattering effects. Such targets are also much easier to prepare and require no replenishment, as is the case with lower \(Z \), gaseous, or liquid targets.

The beam energy is chosen to optimize the signal-to-background ratio, typically a few hundred MeV above the meson production threshold. The exact value depends on whether a proton or a pion beam is used. The two beam configurations currently considered for REDTOP are discussed in the following subsections, where their expected performance and experimental implications are compared.

\subsection{Proton Beam}\label{protonbeam}
The first configurations considered for the REDTOP experiment are based on the use of a high-intensity proton beam impinging on a nuclear target. Such a configuration provides an efficient mechanism for the production of $\eta$ and $\eta^{\prime}$ mesons through the excitation of intermediate baryonic resonances.
Initial Monte Carlo studies indicated that the REDTOP physics program would benefit most from  using a proton beam with an energy range of 1.8--2.0~GeV for $\eta$ production and 3.5--4.0~GeV for $\eta'$ production. In that energy range, the $\eta$ and $\eta'$ production cross sections increase approximately linearly~\cite{Moskal:2002jm,Bass:2018xmz} with the beam energy, while the QCD cross section slightly decreases~\cite{Donnachie:1992ny}. The above choices correspond to beam energies a few hundred MeV above the corresponding meson production threshold, while they are low enough to guaranty that the background affects only modestly the sensitivity to New Physics. A minimum beam intensity of $10^{11}$ POT/s, corresponding to an annual integrated flux of $10^{18}$ POT, is necessary to achieve the desired sensitivity. A continuous-wave 
beam, or one slowly extracted from an accumulation ring, is also a key requirement for REDTOP, as it reduces stress on the trigger system and lowers event pile-up in the detector.

\begin{sloppypar}
Along with the beam properties, the target configuration must be optimized to balance meson production efficiency and detector performance.
An optimal solution is provided by a segmented lithium or beryllium target composed of multiple thin disks. Each disk corresponds to about $2\times 10^{-2}$ interaction lengths (approximately 7.7~mm or 2.3~mm, respectively) and the disks are separated by a few centimeters along the beam axis. Such a configuration enhances meson production while limiting multiple scattering and reducing the probability of secondary-particle interactions, including the $\eta(')$ decay products in neighboring target elements. Under these operating conditions, the expected annual yields are approximately $3.3 \times 10^{13}$ $\eta$ mesons and $5 \times 10^{11}$ $\eta'$ mesons. These statistics are sufficient to explore the vast majority of the physics channels discussed in Sec.~\ref{PhysicsSensitivity}. Furthermore, because only about 1\% of the incident beam is absorbed in the target, active cooling is not required, significantly simplifying the engineering and infrastructure needs of the experiment.

A proton beam with the above specifications is readily available at several high-energy physics and nuclear laboratories around the world. In particular, the accelerator complex at Fermilab~\cite{Lebedev:2017vnu} is especially well suited to meet REDTOP requirements. The Delivery Ring can be operated as an accumulation ring at the beam energies specified above. Of the three available extraction points, only one is currently in use. An experimental hall, previously dedicated to antiproton experiments, is now vacant and is already equipped with the necessary infrastructure to host the REDTOP detector. CERN is also capable of providing a slowly extracted beam from the Proton Synchrotron (PS), although the available intensity is slightly lower than that at Fermilab. Additionally, the new HIAF (High-Intensity heavy-ion Accelerator Facility) laboratory in China meets all the criteria for a super-$\eta$/$\eta'$ factory. The HHAS spectrometer is currently being designed specifically for use at that facility~\cite{Chen:2024wad}.
\end{sloppypar}

\subsection{Pion Beam}\label{pionbeam}
\begin{sloppypar}
An alternative configuration considered for the REDTOP experiment is based on the use of a high-intensity pion beam. A pion beam with kinetic energies in the range of 0.83--1.5 GeV and an intensity comparable to that of the proton option ($10^{11}$ particles per second) would yield a sample four times larger of $\eta/\eta^{\prime}$ mesons. Moreover, it offers a significant advantage, as the QCD background is reduced by approximately 10\%. Additionally, the topology of background events produced by pion-induced interactions is easier to discriminate from signal events compared to those produced by protons. 
More specifically, the charged-particle multiplicity is lower, and there are fewer neutrons that can be misidentified as photons. The momentum of the charged particle is also lower, and they would easily be discriminated by the Čerenkov-TOF. As a consequence, a background event has a far lower probability of triggering the DAQ.
This results in an overall improvement in the signal-to-noise ratio of nearly one order of magnitude.

The experimental conditions can be further tuned by selecting the charge of the pion beam. In particular, when using a $\pi^{-}$ beam on a Li, Be, or $^3\mathrm{He}$ target, the $\eta/\eta^{\prime}$ mesons are predominantly produced in association with a neutron (or triton in the case of $^3\mathrm{He}$). If a dedicated tagging detector is employed to identify the associated neutron or triton, the $\eta/\eta^{\prime}$ mesons can be fully identified, leading to further improvement in the signal-to-background ratio. Conversely, when using a $\pi^{+}$ beam on an LDe target, the $\eta/\eta^{\prime}$ mesons are predominantly produced in association with two protons. In this case, tagging can still be achieved by detecting both nucleons, although  with lower efficiency. In summary, operating REDTOP with a pion beam is expected  to improve the sensitivity to Light Dark Matter (LDM) and other BSM signatures  by approximately a factor of $\sqrt{10}$.

However, implementing a pion beamline is technically more complex and resource intensive than a proton beamline. 
It requires an intermediate production target, where pions are generated by a primary proton beam, together with a dedicated pion collection and focusing system. 
Since these systems typically operate with limited efficiency, the losses in pion production and transport must be compensated for by a sufficiently intense primary proton beam with energy above $1.3\,\mathrm{GeV}$ and a power on the order of 100~kW.

Facilities capable of providing a pion beam with the characteristics and specifications mentioned above are very few in the world.
At the European Spallation Source  (ESS) in Sweden, the planned 1.8--2.0 GeV proton beam could serve as the primary source generating a pion beam with energies around 830~MeV, as required by REDTOP. The produced pions could then be focused using either a van der Meer horn or a direct current magnetic focusing system~\cite{Koetke:1996gk,LosAlamos_Simion}, and subsequently directed onto a secondary target to produce $\eta$ mesons.
A similar possibility exists at the Oak Ridge National Laboratory (USA), where a 1.3~GeV proton beam of the Spallation Neutron Source (SNS) is close to the threshold required to produce an $\approx830$ MeV pion beam. Although the lower proton-beam energy reduces the pion production efficiency, the very high beam intensity available at SNS could largely compensate for this limitation. It should be noted that, at these available lower $\pi^-$ beam energies, only $\eta$ mesons can be produced, because the $\eta^{\prime}$ is kinematically forbidden, since the threshold is $T_\pi \approx 1.30$~GeV.

In summary, both proton and pion beam configurations offer viable paths for the realization of the REDTOP physics program. While proton beams provide a technically simpler solution, pion beams offer improved signal-to-background conditions and enhanced sensitivity to several BSM scenarios. The final choice will depend on the optimization of beam performance, background conditions, and the availability of accelerator facilities capable of hosting and operating a detector of the scale required for REDTOP.
\end{sloppypar}

\section{Roadmap for REDTOP Project\label{par:Timeline-costing}}
\begin{sloppypar}
The REDTOP Collaboration was established in 2015 and currently includes 132 members from 62 institutions. Several groups are actively involved in physics studies, detector simulations, and detector R\&D. As a result, the experimental design is under continuous development to improve sensitivity to BSM physics and to optimize cost and performance. Several accelerator laboratories worldwide are currently being considered as potential hosts for the REDTOP experiment, provided that the beam conditions meet the requirements outlined in Sec.~\ref{sec:beam}. Presently, several laboratories worldwide offer proton beams capable of achieving the necessary yield of more than $>10^{13}~\eta/\text{yr}$. Among them are Fermilab, CERN, J-PARC, ORNL, BNL, ESS, GSI, and HIAF. However, only ORNL and ESS could offer pion beams with characteristics suitable for the REDTOP physics program.

The REDTOP detector design has reached an advanced stage of optimization, supported by a nearly complete simulation and reconstruction framework. While a finalized cost breakdown awaits further technical design, a preliminary cost has been estimated. The current cost estimate includes a 100\% contingency but excludes labor, as presented in Table~\ref{tab:REDTOP-cost-estimate}. Costs for electronics and sensors (e.g., SiPMs, ASICs, LGADs) are based on recent vendor quotes. For major subsystems that use technologies in common with existing or upgraded detectors, cost estimates are scaled appropriately, with adjustments for inflation.
\begin{table}
\centering
\renewcommand{\arraystretch}{1.2}
\begin{tabular}{cc}
\toprule 
{{Component}} & {{M\$}}\\
\midrule
Target+beam pipe & 0.1\\ 
Vertex detector & 2.1\\ 
LGAD tracker & 22.5\\ 
Calorimeter & 22.5\\ 
Čerenkov-TOF & 0.75\\ 
Solenoid & 0.3\\ 
Supporting structure  & 1.3\\ 
Hardware trigger & 2.4\\ 
DAQ+L2 trigger & 1.1\\
\midrule 
Computing & 0.4\\ 
Contingency 100\%  & 53.5\\
\midrule
{Total REDTOP}  & {107}\\
\bottomrule 
\end{tabular}
\caption{Cost estimate for the construction of the baseline REDTOP 
detector, in millions of US dollars (M\$~Y2023). Costs are broken 
down by sub-system, with a 100\% contingency reserve applied to the 
sum of all hardware items.  The LGAD tracker and the ADRIANO2(3) 
calorimeter together account for about 85\% of the detector hardware 
cost, while the total projected investment amounts to approximately 
107~M\$.}
\label{tab:REDTOP-cost-estimate} 
\end{table}

\begin{figure*}[t]
  \centering
\includegraphics[scale=0.4]{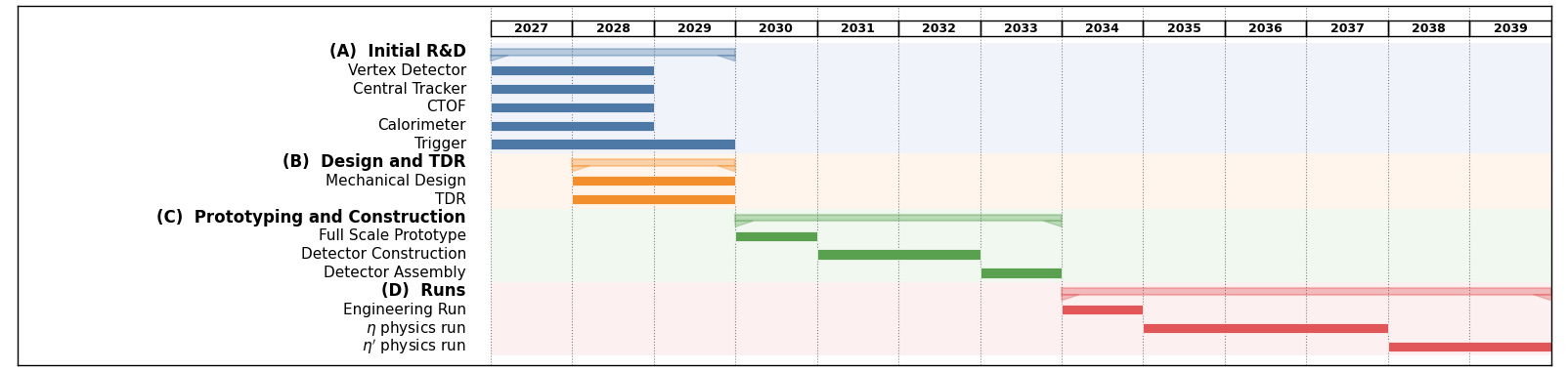} 
\caption{Projected timeline for the REDTOP experiment, spanning 2027-2039. Four overlapping phases are shown: 
\textbf{(A)} initial R\&D on the main detector sub-systems (vertex detector, LGAD tracker, Čerenkov-TOF, calorimeter, and trigger),
\textbf{(B)} mechanical design and preparation of the Technical Design Report (TDR),
\textbf{(C)} prototyping, construction, and on-site assembly, and 
\textbf{(D)} engineering and physics runs, with the dedicated $\eta$ and $\eta'$ campaigns extending through 2039.}
\label{fig:redtop-timeline}
\end{figure*}
Most of the REDTOP detector technologies are already under active development and require only moderate additional R\&D to reach technological readiness. Nevertheless, completing the technical design, engineering, and full integration of the experiment will demand coordinated effort and dedicated support across all participating institutions. Once the experiment is approved at a host laboratory, a staged timeline can be envisioned. The first stage comprises the R\&D and construction period. Completing the detector specific R\&D program is expected to require approximately two to three years, followed by about one year devoted to the detailed design and integration of the full experimental apparatus. The construction, assembly, and installation of the detector infrastructure would then proceed over an additional two-year period. After installation, REDTOP will enter a commissioning phase lasting roughly one year. During this time, engineering runs, calibration campaigns, and system-level performance validation will be carried out. The subsequent data taking period near the $\eta$ production threshold is expected to extend over three years, assuming a beam intensity of approximately $10^{18}$ POT per year. Should the available beam intensity be lower, the data taking period may need to be correspondingly extended unless reduced physics sensitivity is acceptable. Finally, an additional year is reserved as a contingency to accommodate potential delays arising from technical issues or beam availability. Upon completion of the  $\eta$ physics program, a second physics run at a beam energy of 4-5 GeV may be considered to study $\eta^{\prime}$ decays and related physics. A schematic overview of the proposed project timeline, assuming a start date of January 1, 2027,  is shown in Fig.~\ref{fig:redtop-timeline}. 
\end{sloppypar}

\section{Conclusions\label{par:Conclusions}}
\begin{sloppypar}
The REDTOP experiment represents a unique and timely opportunity to probe fundamental questions at the intensity frontier, particularly in the areas of Light Dark Matter, discrete symmetry violations, and advanced detector technologies. Focused on studying rare and suppressed decays of the $\eta$ and $\eta^{\prime}$ mesons, REDTOP is optimized to explore a wide range of BSM scenarios with unprecedented sensitivity and to probe low-energy QCD limits. 
A central pillar of the REDTOP physics program is the search for the particle physics of the dark sector in the MeV--GeV mass range, a region less explored by current experiments. REDTOP is among the few experiments worldwide capable of probing all four theoretical portals connecting the SM to the dark sector—vector, scalar, pseudoscalar, and neutrino, with sensitivities competitive with or surpassing collider-based searches. Its projected ability to explore thermalized LDM with couplings down to $10^{-8}$ makes REDTOP a complementary tool to both direct detection experiments and high-energy colliders.

Equally compelling is REDTOP potential to uncover new sources of $CP$ violation. The experiment will enable high-precision measurements of decay asymmetries, lepton polarization, and angular correlations in processes where SM contributions are suppressed or absent. In particular, REDTOP focuses on flavor-conserving channels and novel observables, such as transverse and longitudinal muon polarization, and opens new ways to detect $CP$-violating effects beyond those accessible via EDM measurements or flavor-changing processes, offering fresh insights into the longstanding puzzle of the matter--antimatter asymmetry in the Universe.

REDTOP also serves as a platform for detector innovation. It integrates cutting edge technologies, including high-granularity, triple-readout ADRIANO3 calorimetry, LGAD-based tracking, and muon polarimetry technologies that not only meet the stringent demands of REDTOP's physics goals but are also directly relevant to future experiments, such as Higgs factories or next-generation precision muon programs. With modest beam power requirements and compatibility with existing accelerator infrastructure, REDTOP is a cost effective and scalable experiment. Its program is designed for long-term operation, with the potential for upgrades and extensions over time. The full realization and scientific exploitation of REDTOP is projected to span just over a decade. However, future upgrades and continued operation could extend the program well beyond that time frame, offering sustained opportunities for discovery and innovation in intensity frontier physics.

In summary, REDTOP is uniquely positioned to explore the nature of dark matter, BSM, $CP$ violation, and nonperturbative QCD in the MeV--GeV energy range. Its targeted approach and innovative design offer both immediate discovery potential and a robust, long-term scientific program.
\end{sloppypar}

\section*{Acknowledgments}
We thank the Open Science Grid project for providing most of the computing and data storage resources used to perform the physics and detector performance studies presented in this article.\\
This manuscript has been authored by Fermi Forward Discovery Group, LLC under Contract No. 89243024CSC000002 with the U.S. Department of Energy, Office of Science, Office of High Energy Physics.
The work of WA is partially supported by the Science, Technology and Innovation Funding Authority (STDF), Egypt, under Grant No. 50806.
%%%%%%%%%%%%%%%%%%%%%%%%%%%%%%%%%%%%%%%%%%%%
\bibliographystyle{utphysmod}
\bibliography{sn-bibliography}
%%%%%%%%%%%%%%%%%%%%%%%%%%%%%%%%%%%%%%%%%%%%
\end{document}